\documentclass[12pt]{article}
\usepackage{epsf}
\usepackage{amsmath,amssymb}
\usepackage[dvips]{graphicx}
\usepackage{subcaption}
\usepackage{float}
\usepackage{comment}
\usepackage{bm}
\begin{document}

\newcommand{\lsim}{\stackrel{<}{_\sim}}
\newcommand{\gsim}{\stackrel{>}{_\sim}}
\newcommand{\ra}{\rightarrow}
\newcommand{\nn}{\nonumber}
\newcommand{\rem}[1]{{$\spadesuit$\bf #1$\spadesuit$}}

\renewcommand{\thefootnote}{\fnsymbol{footnote}}
\setcounter{footnote}{0}

\newcommand{\stoponium}{ \sigma_{\tilde{t}_1} }
\newcommand{\lightstop}{ $\tilde{t}_1$ }

\begin{titlepage}
\def\thefootnote{\fnsymbol{footnote}}
\begin{center}

\hfill UT-16-05\\
\hfill February, 2016\\

\vskip .75in

{\Large\bf
  Di-Higgs Decay of Stoponium at \\ \vspace{0.2em}
  Future Photon-Photon Collider \\
}

\vskip .75in

{\large
  Hayato Ito, Takeo Moroi and Yoshitaro Takaesu
}

\vskip 0.25in

{\em 
Department of Physics, University of Tokyo, Tokyo 113-0033, Japan}

\end{center}

\vskip .5in

\begin{abstract}
We study the detectability of the stoponium
in the di-Higgs decay mode at the photon-photon collider option of the
 International $e^+e^-$ Linear Collider (ILC), whose center-of-mass energy is planned to reach
 $\sim 1$ TeV.
We find that $5\sigma$ detection of the di-Higgs decay mode is possible
 with the integrated electron-beam luminosity of
 $1 {\rm ab}^{-1}$ if the signal cross
section, $\sigma(\gamma \gamma \ra \stoponium \ra hh)$, of ${\cal O}(0.1)$
 fb is realized for the stoponium mass smaller than $\sim$ 800 GeV at 1
 TeV ILC. Such a
 value of the cross section can be realized in the minimal
 supersymmetric standard model (MSSM) with relatively large trilinear
 stop-stop-Higgs coupling constant. 
Implication of the stoponium cross section
measurement for the MSSM stop
 sector is also discussed.
\end{abstract}
\end{titlepage}

\setcounter{page}{1}
\renewcommand{\thefootnote}{\#\arabic{footnote}}
\setcounter{footnote}{0}

%%%%%%%%%%%%%%%%%%%%%%%%%%%%%%%%%%%%%%%%%%%%%%%%%%%%%%%%
\section{Introduction}
\setcounter{equation}{0}
Low energy supersymmetry (SUSY) is an attractive candidate of the
physics beyond the standard model even though the recent LHC
experiment is imposing stringent constraints on the mass scale of
superparticles.  Importantly, there is still a possibility that there
exist superparticles with their masses below TeV scale.  In
particular, a scalar top-quark (stop) with the mass of ${\cal O}(100\ {\rm
  GeV})$ is still allowed if there exists a neutralino whose mass is just
below that of the stop mass; in such a case, even if the stop is
produced at the LHC experiments, its decay products are too soft to be
observed so that it can evade the detection at the LHC.

If there exists stop with its mass of ${\cal O}(100\ {\rm GeV})$, it will
become an important target of future collider experiments.  If the
signal of the stop is discovered at the LHC, the LHC and the
International $e^+e^-$ Linear Collider (ILC) may play important role
to study its basic properties (like the mass and left-right mixing
angle).  It is, however, also important to study the strength of the
stop-stop-Higgs coupling because the Higgs mass in the supersymmetric
model is sensitive to it; measurement of the stop-stop-Higgs coupling
is crucial to understand the origin of the Higgs mass in
supersymmetric model.  It motivates the study of the stop-stop
bound state (so-called stoponium which is denoted as
$\sigma_{\tilde{t}_1}$ in this paper) because decay rate of the
stoponium is crucially depends on such a coupling. If we 
observe the process of $\sigma_{\tilde{t}_1}\rightarrow hh$, we can acquire information about the stop-stop-Higgs coupling.

Photon-photon colliders may be useful to perform such a
study~\cite{Gorbunov:2000tr, Gorbunov:2000nd}.\footnote{For the
stoponium studies at other colliders, see 
%\cite{Herrero:1987df}-- \cite{Batell:2015zla}.}
\cite{stoponium_other_colliders}.}
A photon-photon collider is one of the options of the ILC and can be
realized by converting high-energy electron (or positron) beam of the
ILC to the backscattered high-energy photon.  It has been intensively
discussed that the photon-photon collider can be used to study the
properties of Higgs and other scalar particles~\cite{scalar_at_photon_collider}.
One of the advantages of photon-photon
colliders is that the single production of some scalar particles
(including the Higgs boson) is possible so that the kinematical reach
is close to the total center-of-mass (c.m.) energy; this is a big contrast to
other colliders with $pp$ and $e^+e^-$ collision.  The single
production of the stoponium bound state is also possible with the
photon-photon collision, and hence it is interesting to consider the
stoponium study at the photon-photon collider.

In this paper, we investigate how and how well we can study the property of
the stoponium at the photon-photon collider, paying particular attention
to the process of
$\gamma\gamma\rightarrow\sigma_{\tilde{t}_1}\rightarrow hh$.  We
calculate the cross section of the stoponium production process at the
photon-photon collider. We also estimate backgrounds, and discuss
the possibility of observing the stoponium production process at the
photon-photon collider. Implication of the cross section
measurement of the stoponium di-Higgs decay mode for the MSSM stop
 sector is also discussed.

The organization of this paper is as follows. In Section 2, we discuss
the theoretical framework of the stoponium and its production cross
section and decay widths at the photon-photon collider. The
detectability of the stoponium in the di-Higgs decay mode is investigated
in Section 3. In Section 4 we discuss the implication of the cross
section measurement of the stoponium production and di-Higgs decay for
the MSSM stop sector. Then we provide our summary in Section 5.

\section{Stoponium: basic properties}
\label{sec:stoponium}
\setcounter{equation}{0}

\subsection{Framework}
Let us first summarize the framework of our analysis.  We assume the
minimal supersymmetric standard model (MSSM) as the underlying theory. The relevant part of the superpotential
for our study is
given by
\begin{align}
  W = y_t \epsilon_{ij} \hat{\bar t}_R \hat{Q}_L^i \hat{H}_u^j + 
  \mu \epsilon_{ij} \hat{H}_u^i \hat{H}_d^j,
\end{align}
where $y_t$ is the top Yukawa coupling, $H_u$, $H_d$, $t_R$ and $Q_L$ denote up- and down-type Higgses,
right-handed top quark, and third-generation quark-doublet, respectively,
and ``hat'' is used for the corresponding superfields. In addition, $i$ and $j$ are $SU(2)_L$
indices, while the color indices are omitted for simplicity.
The relevant part of the soft SUSY breaking terms is
\begin{align}
  {\cal L}_{\rm soft} = 
  -m_{\tilde{t}_R}^2 |\tilde{\bar t}_R|^2 - m_{\tilde{Q}_L}^2 |\tilde{Q}_L|^2
  + y_t A_t 
  ( \epsilon_{ij} \tilde{\bar t}_R \tilde{Q}_L^i H_u^j + {\rm h.c.}),
\end{align}
where ``tilde'' is used for superpartners.

Neglecting the effects of flavor mixing, the stop mass terms are expressed as
\begin{align}
{\cal L}_{\rm mass} = -(\tilde{t}_L^*, \tilde{t}_R^*)
  \left(
    \begin{array}{cc}
      m_{\tilde{Q}_L}^2 + m_t^2 +D_L & -m_t X_t
      \\
      -m_t X_t & m_{\tilde{t}_R}^2 + m_t^2 + D_R
    \end{array}
  \right)
\left(
 \begin{array}{c}
  \tilde{t}_L \\
  \tilde{t}_R
 \end{array}\right),
\end{align}
where 
$m_t$ is the top-quark mass, $\tilde{t}_R \equiv \tilde{{\bar t}}_R^*$, 
\begin{align}
  X_t \equiv A_t +\mu \cot\beta,
\label{eq:Xt}
\end{align}
and
\begin{align}
  D_L &\equiv
  m_Z^2 \cos 2\beta 
  \left( \frac{1}{2} - \frac{2}{3} \sin^2 \theta_{\rm W} \right),
  \\
  D_R &\equiv
  m_Z^2 \cos 2\beta 
  \left( \frac{2}{3} \sin^2 \theta_{\rm W} \right),
\end{align}
with $\theta_{\rm W}$ being the Weinberg angle, $m_Z$ the $Z$-boson
mass, and $\tan\beta\equiv\langle H_u^0\rangle/\langle H_d^0\rangle$.
$A_t$ and $\mu$ parameters are taken to be real. The mass eigenstates are given by the linear combination of the left-
and right-handed stops; we define the mixing angle
$\theta_{\tilde{t}}$ as
\begin{align}
  \left(
    \begin{array}{c}
      \tilde{t}_1 \\ \tilde{t}_2
    \end{array}
  \right) = 
  \left(
    \begin{array}{c c}
      \cos \theta_{\tilde{t}} & \sin \theta_{\tilde{t}} \\
      - \sin \theta_{\tilde{t}} & \cos \theta_{\tilde{t}}
    \end{array}
  \right)
  \left(
    \begin{array}{c}
      \tilde{t}_L \\ \tilde{t}_R
    \end{array}
  \right),
\end{align}
where $\tilde{t}_1$ and $\tilde{t}_2$ are lighter and heavier mass
eigenstates with the masses of $m_{\tilde{t}_1}$ and
$m_{\tilde{t}_2}$, respectively. 

Before closing this subsection, we comment on the lightest MSSM Higgs boson
mass. The lightest Higgs mass is sensitive to the masses of stops as
well as to the $A_t$ parameter through radiative corrections.  
With the stop masses being fixed, the lightest
 Higgs mass becomes equal to the observed Higgs mass (which is
taken to be $m_h\simeq 125.7\ {\rm GeV}$ throughout our study) for
four different values of $A_t$; two of them are positive and others
are negative. We call these as positive-large, positive-small,
negative-large, and negative-small solutions of $A_t$, where large and
small solutions correspond to those with large and small values of
$|A_t|$. 

\subsection{Stoponium production at a photon-photon collider and its decay}
\label{sec:production-decay}
Due to the strong interaction, a stop and an anti-stop can form a bound
state, called stoponium. In this analysis, we concentrate on the case where the decay rate of
a stop is negligibly small.  
 Because we are interested in the
collider study of the stoponium, we concentrate on the bound state of
the lighter stop.  The lowest bound state, $\stoponium$, has the quantum number of
$J^{PC}=0^{++}$, and hence its resonance production does not occur at
$e^+e^-$ colliders.  At photon-photon colliders, on the contrary,
the process $\gamma\gamma\rightarrow\stoponium\rightarrow F$
may occur, where $F$ denotes final-state particles of the stoponium
decay. High energy photon-photon collisions can be achieved by photons originating
from backscattered lasers off electron beams, and this possibility was
discussed in detail~\cite{Ginzburg:1981vm,Ginzburg:1982yr}. For a concrete
discussion, we assume a photon-photon collider utilizing an upgraded International Linear Collider (ILC)
whose energy is planned to reach $\sqrt{s_{ee}} = 1\,{\rm
TeV}$ 
~\cite{Behnke:2013xla}. 

With the c.m.\ energy of colliding photons, $\sqrt{s_{\gamma\gamma}}$, being fixed,\footnote{In this article, we denote the center-of-mass energy of colliding photons by $\sqrt{s_{\gamma\gamma}}$ 
and that of colliding electrons and positrons by $\sqrt{s_{ee}}$.} the cross section for the
process $\gamma\gamma\rightarrow\stoponium\rightarrow F$ can be written
as~\cite{Ginzburg:1982yr}
\begin{align}
  \hat{\sigma}
  (\gamma\gamma\rightarrow\stoponium\rightarrow F; s_{\gamma\gamma}) =
  \left( \frac{1+\xi_2\xi_2'}{2} \right)
  \hat{\sigma}_{++}
  (\gamma\gamma\rightarrow\stoponium\rightarrow F; s_{\gamma\gamma}).
\end{align}
Here $\xi_2$ and $\xi_2'$ are the Stokes parameters of the initial-state
photons 
where $\xi_2 =
\pm 1$ corresponds to the photons with helicity $\pm
1$.
In this study we only consider axially symmetric electron beams, and then other components of the Stokes parameters
($\xi_{1,3}$) are negligible~\cite{Ginzburg:1982yr}.
$\hat{\sigma}_{++}$ (= $\hat{\sigma}_{--}$) is the cross section for
photon collisions with circular polarization and given with the
Breight-Wigner approximation by
\begin{align}
  \hat{\sigma}_{\lambda\lambda'}
  (\gamma\gamma\rightarrow\stoponium\rightarrow F; s_{\gamma\gamma}) =
  \frac{16\pi m_\sigma^2}{s_{\gamma\gamma}}
  \frac{\Gamma (\stoponium\rightarrow \gamma \gamma)
    \Gamma (\stoponium\rightarrow F)}
  {(s_{\gamma\gamma}-m_\sigma^2)^2 + m_\sigma^2 \Gamma_\sigma^2}
  \delta_{\lambda\lambda'} ,
\end{align}
where $\Gamma_\sigma$ and $\Gamma (\sigma_{\tilde{t}_1}\rightarrow
\gamma\gamma/F)$ are the total and partial decay widths of the stoponium, respectively, $\lambda=\pm$ and $\lambda'=\pm$
are polarizations of initial-state photons, and $m_{\sigma}$ is the mass
of the stoponium, 
which is roughly estimated as 
\begin{equation}
 m_{\sigma} = 2m_{\tilde{t}_1},
\end{equation}
throughout this paper. 
Detailed calculations in Ref.~\cite{Hagiwara:1990sq} shows that the
error of this estimation is $\sim 0.5\%$ and negligible for our discussion.

Since backscattered photons off electron beams are not monochromatic,
the cross section at the photon-photon collider is given by\footnote{In
this article, we define cross sections for the photon-photon collider as the number of events normalized by the luminosity of electron beams, $L_{ee}$.}
\begin{align}
  \sigma
  (\gamma\gamma\rightarrow\sigma_{\tilde{t}_1}\rightarrow F; s_{ee}) 
  &= 
  \frac{1}{L_{ee}} 
  \int_0^{y_m} dy dy'
  \frac{d^2L_{\gamma\gamma}}{dydy'}
  \hat{\sigma}
  (\gamma\gamma\rightarrow\sigma_{\tilde{t}_1}\rightarrow F; s_{\gamma\gamma} =
 yy's_{ee}),
\end{align}
using the
luminosity function of backscattered photons~\cite{Ginzburg:1981vm,Ginzburg:1982yr,Ginzburg:1999wz} denoted by $d^2L_{\gamma\gamma}/dydy'$
(with $y$ and $y'$ being the photon energies normalized by the energy
of the electron beam $E_e$ in the laboratory frame).\footnote{For more details
about the luminosity function, see Appendix \ref{app:luminosityfn}}
Here, $y_m \equiv x/(x+1), \,x\equiv 4E_e\,\omega_0/m_e^2$ with $\omega_0$
being the averaged energy of the laser photons in a laboratory frame,
and $L_{ee}$ and $s_{ee} (= 4E_e^2)$ are the luminosity and c.m.\ energy of the electron beams, respectively. 
We take $x = 4.8$ to maximize $y_m$ without spoiling
the photon luminosity~\cite{Ginzburg:1999wz}.
Since we consider the case where $\Gamma_\sigma \sim {\cal O}(10^{-3})\,\, {\rm GeV} \ll
m_\sigma$, we use the narrow-width
approximation and obtain
\begin{align}
  \sigma
  (\gamma\gamma\rightarrow\sigma_{\tilde{t}_1}\rightarrow F;s_{ee}) 
  & \simeq
  \frac{16\pi^2\Gamma(\sigma_{\tilde{t}_1}\rightarrow \gamma\gamma)}
  {m_\sigma}
  {\rm Br} (\sigma_{\tilde{t}_1}\rightarrow F) \nonumber\\
 & \times \frac{1}{s_{ee}L_{ee}} \int_{z_0^2/y_m}^{y_m} \frac{dy}{y}
  \left[\frac{d^2L_{\gamma\gamma}}{dydy'}
    \frac{1 \pm \xi_2(y) \xi_2 (y')}{2} \right]_{y'= z_0^2/y},
  \label{sigma(pp)}
\end{align}
with $z_0 = m_\sigma / \sqrt{s_{ee}}$.
The $y$ dependence of $\xi_2$ is given by
\cite{Ginzburg:1981vm,Ginzburg:1982yr}
\begin{align}
  \xi_2(y) &= \frac{C_{20}(x,y)}{C_{00}(x,y)},
  \label{C20}
\end{align}
where 
\begin{align}
 C_{00}(x,y) &=  \frac{1}{1-y} - y +( 2r-1 )^2 - \lambda_e P_l \,x \,r ( 2r-1 )( 2-y ),
  \label{C00}
  \\
  C_{20}(x,y) &= \lambda_e r x \left[ 1 + ( 1 - y )( 2r -1 )^2 \right] -P_l\,( 2r - 1 )\left( \frac{1}{1-y} + 1-y \right),
\end{align}
with $\lambda_e/2$ and $P_l$ being the mean helicities of initial
electrons and laser photons, respectively, and $r= y /x (1-y)$.
In our numerical calculation, we take $\lambda_e=0.85$
and $P_l=-1$. 

The decay rates of the stoponium are related to the matrix elements for
the pair-annihilation processes of the stop and anti-stop.  For the case
of two-body final states, i.e., $\sigma_{\tilde{t}_1}\rightarrow ff'$,
the decay width is related to the matrix element of the scattering
process ${\cal M}(\tilde{t}_1\tilde{t}_1^*\rightarrow ff')$ as \cite{Drees:1993uw}
\begin{align}
  \Gamma (\sigma_{\tilde{t}_1}\rightarrow ff') &= 
  \frac{3}{32\pi^2 (1+\delta_{ff'})} \beta_{ff'}
  \frac{|R_{1S}(0)|^2}{m_\sigma^2}
  \sum_{\rm spin,\,color} 
  |{\cal M}(\tilde{t}_1\tilde{t}_1^*\rightarrow ff')|_{v\rightarrow 0}^2,
\end{align}
where
\begin{align}
\beta_{ff'}^2 &= \left( 1 -\frac{m_f^2}{m_\sigma^2}
 -\frac{m_{f'}^2}{m_\sigma^2} \right)^2 -4\,\frac{m_f^2}{m_\sigma^2}\frac{m_{f'}^2}{m_\sigma^2},
\end{align}
$v$ is the velocity of the stops in the initial state,
 and $R_{1S}(0)$ is the stoponium radial wave
function at $r=0$ (with $r$ being the distance between $\tilde{t}_1$ and
$\tilde{t}_1^*$).  In our study, we use \cite{Hagiwara:1990sq, Drees:1993uw, Martin:2008sv}
\begin{align}
  \frac{|R_{1S}(0)|^2}{m_\sigma^2} = 0.1290 + 0.0754L + 0.0199 L^2
 +0.0010 L^3 \,[{\rm GeV}],
\end{align}
where $L=\ln (m_{\tilde{t}_1}/250 \, {\rm GeV})$.  The matrix elements
for the stop pair-annihilation
processes relevant to our study are summarized in Appendix
\ref{app:melements}. As we have mentioned, we consider the case where the
decay width of the stop is much smaller than the total decay width of
the stoponium. 
This is the case when the mass difference between the lighter stop and
the lightest supersymmetric particle (LSP), which is assumed to be the
lightest neutralino in our analysis, is small enough.
\begin{figure}[t]
 \centering
  \includegraphics[width=0.45\columnwidth]{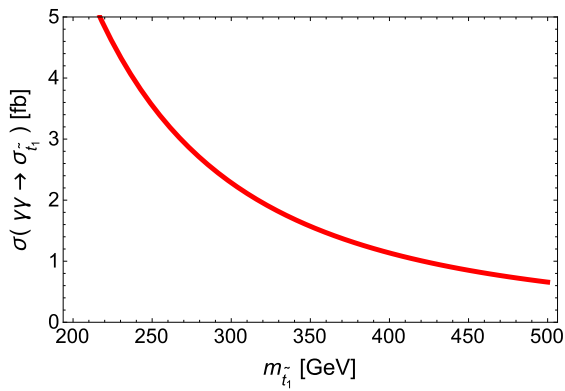}
  \caption{\small The stoponium production cross section $\sigma
    (\gamma\gamma\rightarrow\sigma_{\tilde{t}_1})$ as a function of
    the lightest stop mass $m_{\tilde{t}_1}$.  The center-of-mass
    energy of the electron beams is taken to be $\sqrt{s_{ee}}=1.26 \,m_\sigma$, which
    maximizes the cross section.}
  \label{fig:cstotal}
\end{figure}

In Fig.\ \ref{fig:cstotal}, we plot the stoponium
production cross section 
\begin{align}
  \sigma
  (\gamma\gamma\rightarrow\sigma_{\tilde{t}_1}) 
  = 
  \frac{16\pi^2\Gamma(\sigma_{\tilde{t}_1}\rightarrow \gamma\gamma)}
  {m_\sigma}
\frac{1}{s_{ee}L_{ee}} \int_{z_0^2/y_m}^{y_m} \frac{dy}{y}
  \left[\frac{d^2L_{\gamma\gamma}}{dydy'}
    \frac{1 \pm \xi_2(y) \xi_2 (y')}{2} \right]_{y'= z_0^2/y},
  \label{sigma_tot}
\end{align}
taking $\sqrt{s_{ee}} = 1.26\,m_\sigma$, which maximizes the cross
section.
The stoponium production cross section can
be as large as $O(1)$ fb for $m_{\tilde{t}_1}\lesssim 500\ {\rm
  GeV}$.  

\begin{figure}[t]
 \centering
  \includegraphics[width=0.43\columnwidth]{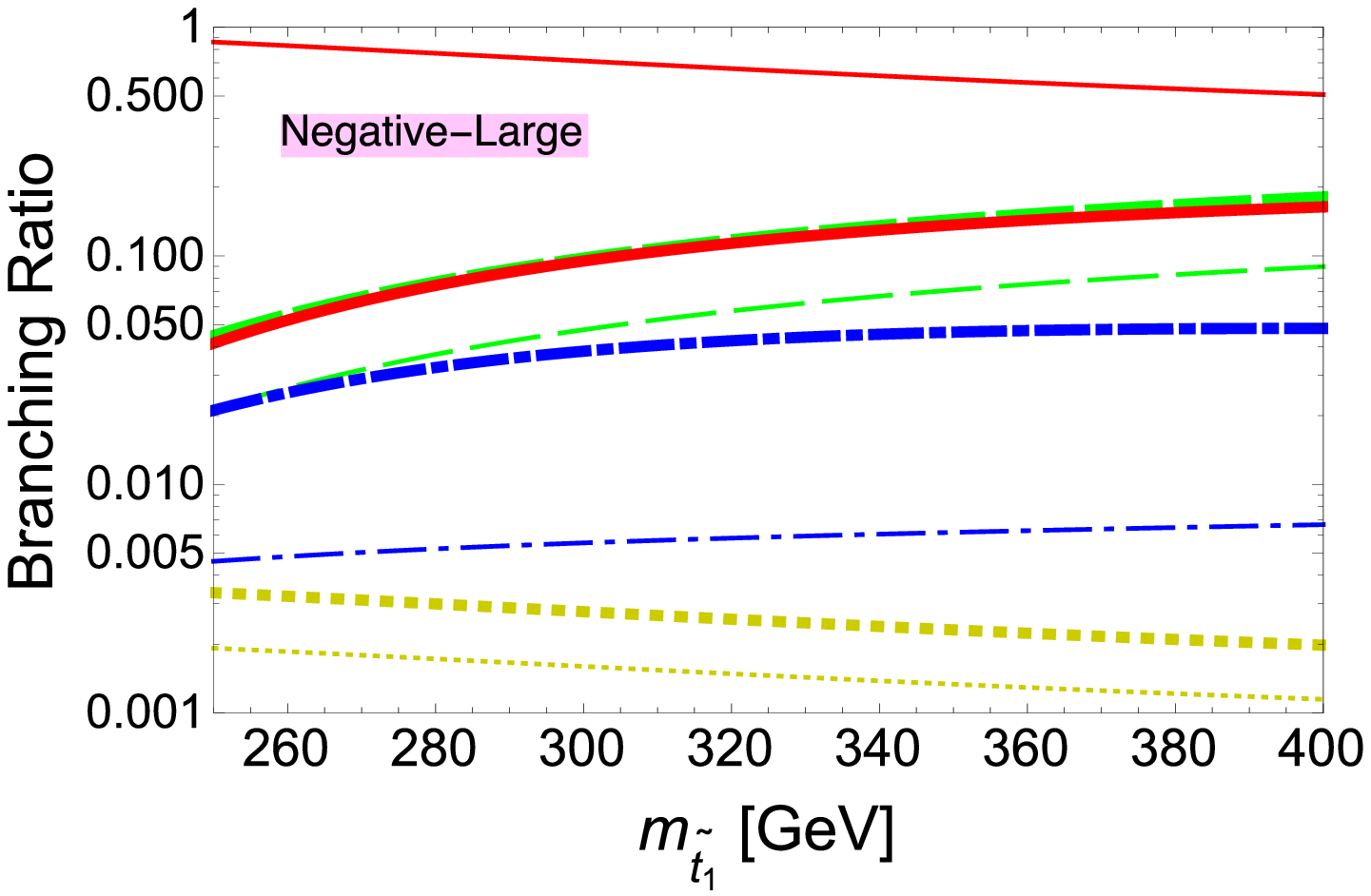}
  \includegraphics[width=0.49\columnwidth]{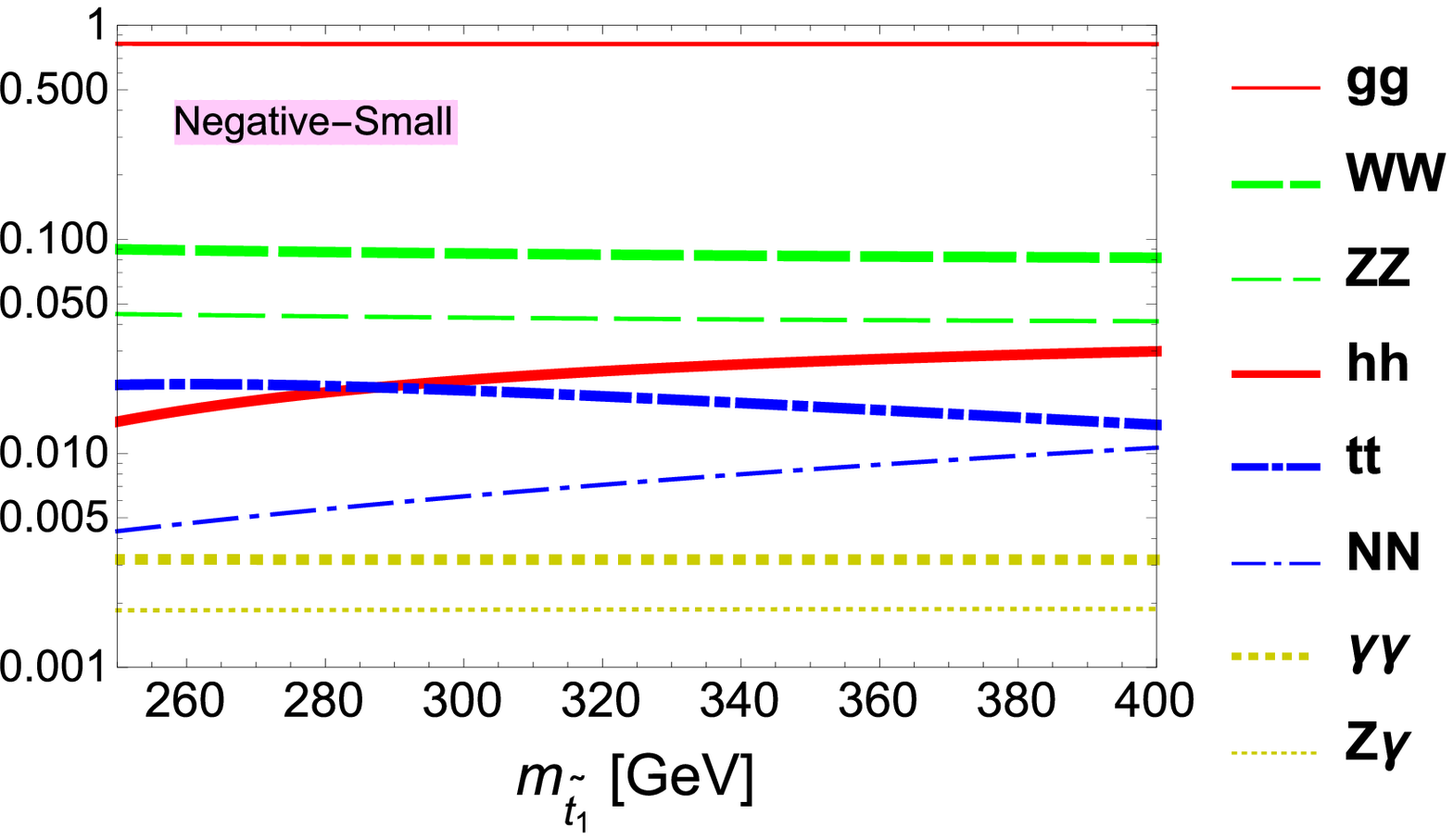}
 \caption{ The branching ratios of the stoponium as functions
    of the lightest stop mass, taking $m_{\tilde{t}_2}=4\ {\rm TeV}$, $m_{\chi^0_1} =
m_{\tilde{t}_1} -50\,{\rm GeV}$, $\tan\beta=10$, $\mu=2\ {\rm TeV}$ with negative-large (left) and
    negative-small (right) solutions of the $A_t$, which give
    $m_h=125.7\ {\rm GeV}$. In the legend, $NN$ stands for the branching
 ratio of the decay to the lightest neutralino pair.}
  \label{fig:br}
\end{figure}
In Fig.\ \ref{fig:br}, we show the branching ratios of the stoponium as
functions of the lightest stop mass, taking
$m_{\tilde{t}_2}=4\ {\rm TeV}$, $m_{\chi^0_1} =
m_{\tilde{t}_1} -50\,{\rm GeV}$, $\tan\beta=10$, $\mu=2\ {\rm TeV}$ and negative-large
or negative-small solutions of the $A_t$ parameter.  
The positive $A_t$ solutions give similar branching ratios. 
The $gg$ decay mode dominates the stoponium decay, and this may be an useful
mode for stoponium searches at photon-photon colliders. However, the
$gg$ branching ratio is completely determined by the strong coupling constant, and measuring this mode does not give much information on SUSY interactions. Therefore, we do
not investigate this mode in this study. Although $WW$ and $ZZ$ decay modes have
non-negligible branching ratios of ${\cal O}(1-10)\%$, they suffer from large SM
backgrounds~\cite{Gorbunov:2000tr, Gorbunov:2000nd, Gounaris:1999hb}. 
We therefor investigate $hh$ decay mode in the
following sections as a probe to the
SUSY interactions, especially to the stop sector. 
We will see in the next section that the signal-to-background ratio
of the $\gamma\gamma\rightarrow\stoponium\rightarrow hh$ process may be
large enough to be observed at the ILC-based photon-photon collider.

%%%%%%%%%%%%%%%%%%%%%%%%%%%%%%%%%%%%
\section{$\stoponium \rightarrow hh$ search at a photon-photon collider}
\label{section:collider_study}
In this section, we discuss a search strategy for the $hh$ decay mode of
the stoponium
at the photon-photon collider.
We assume an upgraded
ILC, whose energy is planned to reach
$\sqrt{s_{ee}} = 1 \rm{TeV}$~\cite{Behnke:2013xla} with the integrated
luminosity of $\mathcal{L}_{ee} \sim 1\,{\rm ab}^{-1}$. 

For our numerical calculation, we adopt four sample SUSY model points
with $m_{\tilde{t}_1}=250$, $300$, $350$, and $400\ {\rm GeV}$ as
summarized in Table~\ref{table:samplepoints}. 
\begin{table}[t]
  \begin{center}
    \begin{tabular}{lrrrr}
      \hline\hline
      & Point 1 & Point 2 & Point 3 & Point 4
      \\
      \hline
      $m_{\tilde{t}_1}$ & 250 & 300 & 350 & 400
      \\
      $m_{\tilde{t}_2}$ & 3480 & 3810 & 4110 & 4080
      \\
      $A_t$ & -4370 &-4940 & -5460 & -5670
      \\
      $m_{\tilde{\chi}_1^0}$ & 150 & 250 & 300 & 350
		           \vspace{0.3em}
      \\
      \hline
      $\sqrt{s_{ee}}$ [GeV] & 625 & 750 & 875 & 1000
      \\
      $\sigma( \gamma \gamma \rightarrow \stoponium \rightarrow hh )$ [fb] & 0.34 & 0.26 & 0.2 & 0.18
      \\
      \hline\hline
    \end{tabular}
    \caption{The sample SUSY models we adopt in our collider analyses.
All the SUSY parameters are given in units of GeV.
    We take $\tan\beta = 10$ and $\mu = 2$ TeV in all the sample points. Other SUSY particle
   masses and soft-breaking trilinear couplings are taken to be $ 2$ TeV. 
   These sample points realize $m_h = 125.7$ GeV. The
   employed c.m.\ energy of the electron beams and the cross section of the $\gamma \gamma \rightarrow
   \stoponium \rightarrow hh$ process are also shown for each sample point.}
    \label{table:samplepoints}
  \end{center}
\end{table}
We assume that the lighter stop and the lightest neutralino,
$\tilde{\chi}^0_1$, will be discovered before the photon-photon collider
experiment is carried out. In addition, if the stop is within the kinematical reach
of the photon-photon collider, detailed studies of the stop will have
been already performed with the $e^+ e^-$ collisions at the ILC, and hence we
 also assume that the basic properties of the lighter stop such as the
 mass and the chirality will be measured at the ILC
before the start of the photon-photon collider.  
We consider the cases where the bino is the lightest SUSY particle (LSP) with $m_{\tilde{\chi}_1^0}=150, 250, 300$, and $350$ GeV 
for each $m_{\tilde{t}_1}$, respectively.\footnote{We don't consider the relic
abundance of the LSP in this study.}
Other SUSY particles are assumed to be sufficiently heavy ($\gsim 2$ TeV) and
irrelevant to our photon collider study.  
$\tan\beta$ and $\mu$ parameter are taken to be 10 and 2 TeV,
respectively, for all the sample points, and the Higgs mass of 125.7 GeV is realized by 
adjusting the $A_t$ parameter and the heavier stop mass,
$m_{\tilde{t}_2}$. (For our numerical calculation of the Higgs mass, we
use
{\tt FeynHiggs \!v2.11.3}~\cite{FeynHiggs}.)
%\cite{Heinemeyer:1998yj}--\cite{Hahn:2013ria}.)
In order to maximize the stoponium
production cross section, we adjust the c.m.\ energy of the electron
beams as $ \sqrt{s_{ee}} \sim 1.26\,m_\sigma $
for each sample point shown in Table~\ref{table:samplepoints}. The
cross sections of the process $\gamma \gamma \ra \stoponium \ra hh$ for
those sample points are also shown in the Table.

\subsection{Signal event selection}
From Table~\ref{table:samplepoints} 
we see that the cross sections $\sigma(\gamma\gamma \rightarrow \stoponium
\rightarrow hh)$ for the sample points are of ${\cal O}$(0.1)\,fb, and then the
process would give only ${\cal O}$(100) events at $1\,{\rm ab}^{-1}$. Therefore, 
we use the main $b\bar{b}$ decay mode of the Higgs boson for the signal
process, i.e., $\gamma\gamma \rightarrow \stoponium
\rightarrow hh \rightarrow b\bar{b}b\bar{b}$.

In order to simulate the signal process, 
we generate events where a scalar particle (which corresponds to the stoponium) is produced at the
photon-photon collider and decayed to a
Higgs pair with their subsequent decays to $b\bar{b}$, using {\tt
MadGraph5\_aMC@NLO\,v2}~\cite{Alwall:2014hca}; the luminosity function
of the colliding photons~\cite{Ginzburg:1982yr,Ginzburg:1999wz} are
implemented by modifying the electron PDF routines in {\tt
MadGraph5}.  
The cross section of the events is normalized to that of the signal
process according to Eq.\ \eqref{sigma(pp)}.
The generated events are then showered with {\tt PYTHIA\,v6.4} \cite{Sjostrand:2006za} and passed to {\tt DELPHES\,v3} \cite{deFavereau:2013fsa} for fast detector simulations.
In the detector simulation, we assume energy resolutions of 2\%/$\sqrt{E \textrm{(GeV)}}
\oplus 0.5$\% and 50\%/$\sqrt{E \textrm{(GeV)}} \oplus 3$\%
for an electromagnetic calorimeter
and hadron calorimeter, respectively, based on ILC TDR~\cite{Behnke:2013lya}.
{\tt FastJet\,v3} \cite{Cacciari:2011ma} is employed for jet clustering
using the anti-$k_t$ algorithm~\cite{Cacciari:2008gp} with the distance
parameter of 0.5.

From the generated events, we first select events containing more than four jets and satisfying ${p_{T}} > 30$ GeV and $| \eta | < 2.0$
for all of  the four highest $p_T$ jets (Preselection), where $p_T$ and
$\eta$ are the transverse momentum and pseudo-rapidity, respectively.  
We then impose the following cuts successively:
\begin{align*}
 {\rm S1}:&\, m_{\sigma} -60\,{\rm GeV} \le M_{\rm{4jets}} \le m_{\sigma} +40\,{\rm GeV}.\\
 {\rm S2}:&\, N_{\rm{b\mathchar`-tag}} \ge 3.\\
 {\rm S3}:&\, 105.7\,{\rm GeV} \le M_1 \le 130.7\,{\rm GeV},\\
          &\, 100.7\,(105.7)\,{\rm GeV} \le M_2 \le 130.7\,{\rm GeV
 \hspace{1em} (for \,the \,Point \,1 \,(2, \,3, \,4))}.\\
 {\rm S4}:&\, \min\{ \Delta R_1 , \Delta R_2 \} \le 1.4, \\
          &\, \max\{ \Delta R_1 , \Delta R_2 \} \le 1.8.
\end{align*}
Here $M_{\rm{4jets}}$ is the invariant mass of the four highest $p_T$ jets.
$N_{\rm{b\mathchar`-tag}}$ is the number of b-tagged jets in each event, where 
we assume 80\% b-tag efficiency, and 10\% and 0.1\% mis-tag rates
for $c$ jets and $u, d, s$ jets, respectively.
In S3 and S4, $M_{1(2)}$ and $\Delta R_{1(2)}$ are defined as
follows. We first divide the leading four jets into two jet pairs. Among
three possible pairings, we choose the one which minimizes 
$(M_1-m_h)^2 +(M_2 -m_h)^2$, where $M_1$ and $M_2$ are the invariant
masses of the jet pairs such that 
\begin{equation}
 |M_1 -m_h| < |M_2 -m_h|.
\end{equation}
$\Delta R_{1(2)}$ is defined as
\begin{align}
 \Delta R_{1(2)} = \sqrt{(\Delta \eta_{1(2)})^2 +(\Delta \phi_{1(2)})^2},
\end{align} 
where $\Delta \eta_{1(2)}$ and $\Delta \phi_{1(2)}$ are the differences of
pseudo-rapidities and azimuthal angles between the paired
jets with the invariant mass of $M_{1(2)}$, respectively.

\subsection{Backgrounds}
\label{analysis_hh_zz}
After imposing the selection cuts, the relevant background processes
are the non-resonant $hh$, $bb\bar{b}\bar{b}$, $b\bar{b}c\bar{c}$,
$cc\bar{c}\bar{c}$, $b\bar{b}q\bar{q}$ (where $q = u, d, s$),
$t\bar{t}$, $ZZ$, $W^+W^-$ and $W^+W^-Z$ production processes. 
 The event numbers after all the selection cuts are imposed are estimated for the
 above background processes as in the signal process case, except the
 non-resonant $hh$ and $ZZ$ backgrounds, which are loop induced
 processes.

As for $hh$ and $ZZ$ backgrounds, we use approximate estimations; the event
numbers of the non-resonant $hh$ background after each cuts are
estimated with $\sim$ 15\% uncertainty, and the event numbers of the
 $ZZ$ background are estimated as the upper bounds after all cuts are
 applied. We will see in Sec.~\ref{sec:results} that even these rough estimations are enough for
our study and leave more detailed estimations for future works.

In the following, we describe our procedure to estimate the non-resonant
$hh$ and $ZZ$ backgrounds.
The production cross sections of background processes at the photon-photon
collider can be expressed in the similar way as for the signal
process discussed in Sec.~\ref{sec:production-decay}: 
\begin{align}
  \sigma
  (\gamma\gamma\rightarrow F; s_{ee}) 
  &= 
  \frac{1}{L_{ee}} 
  \int_0^{y_m} dy dy'
  \frac{d^2L_{\gamma\gamma}}{dydy'}
  \hat{\sigma}
  (\gamma\gamma\rightarrow F; s_{\gamma\gamma} = yy's_{ee}) \nn\\[+3pt]
&= \sum_{\lambda\lambda^{'} = ++, +-} \int^{y_m}_0 %\label{eq:xsecBG_0}
 \hspace{-1em}dz 
\left[ \frac{1}{L_{ee}} \frac{d L_{\gamma \gamma}}{dz} \frac{1
 \pm \xi_2 \xi'_2}{2} \right]\!\!\left(z\right)
\,\,\hat{\sigma}_{\lambda \lambda^{'}}
  (\gamma\gamma\rightarrow F; s_{\gamma\gamma} = z^2s_{ee}),
\label{eq:xsecBG}
\end{align}
where 
\begin{align}
\left[ \frac{1}{L_{ee}} \frac{d L_{\gamma \gamma}}{dz} \frac{1 \pm \xi_2 \xi'_2}{2} \right]\!\!\left(z\right)
&\equiv 
2z \int^{y_m}_{z^2 / y_m} \frac{dy}{y} \frac{1}{L_{ee}} \frac{d^2L_{\gamma \gamma}}{dy dy'} 
\frac{1 \pm \xi_2 (y) \xi_2(y')}{2} \Bigl|_{y' = z^2 / y},
\end{align}
with $z = \sqrt{s_{\gamma\gamma}/s_{ee}}$.
In the second line,
 contributions from $\xi_1$ and $\xi_3$ are negligible since we consider axial symmetric
 electron beams; the sign in front of the stokes
 parameters are taken to be positive (negative) for $\lambda \lambda^{'}
 = ++ (+-)$.

\subsubsection*{$\bm{\gamma \gamma \rightarrow hh \rightarrow bb\bar{b}\bar{b}}$}

First, let us discuss the non-resonant Higgs pair production
process. The dominant background contributions are from Higgs
pairs decaying to bottom quarks. 
Based on Eq.~\eqref{eq:xsecBG}, the cross section after all selection cuts are imposed, 
$\sigma_{\rm cut}(\gamma\gamma\rightarrow hh\rightarrow
bb\bar{b}\bar{b}; s_{ee})$, is given by
\begin{align}
\sigma_{\rm cut}(\gamma\gamma\rightarrow hh\rightarrow
 bb\bar{b}\bar{b}) &= {\rm Br}(h \rightarrow b\bar{b})^2 
 \sum_{\lambda\lambda^{'} = ++, +-}
\int^{y_m}_{0} dz \int^{y_m}_{z^2/y_{m}} dy\,\,
\frac{1}{L_{ee}} \frac{d^2 L_{\gamma \gamma}}{dz dy} \left( \frac{1
 \pm \xi_2 \xi'_2}{2} \right) \nonumber\\[+3pt]
&\mspace{30mu} \times \int^1_{0}d\cos\theta^* \frac{d\hat{\sigma}_{\lambda
 \lambda^{'}}\left(\gamma \gamma\rightarrow hh; s_{\gamma\gamma} = z^2 s_{ee}, \theta^* \right)}{d\cos\theta^*} \,\varepsilon_{\lambda \lambda^{'}}(z,y,\theta^*),
\label{eq:sigma_est}
\end{align}
where $\varepsilon_{\lambda \lambda^{'}}(z,y, \theta^*)$ is the 
total efficiency of all the selection cuts for events with a c.m.\ energy
$\sqrt{s_{\gamma\gamma}} = z\sqrt{s_{ee}}$, a total energy measured in the laboratory frame $E_{\rm{lab}}=(y+z^2/y)E_{ee}$,
 and Higgs scattering angle $\theta^*$ in the c.m.\ frame of the
$\gamma\mathchar`-\gamma$ collision. 

\begin{figure}[t]
 \centering
  \includegraphics[width=0.7\columnwidth]{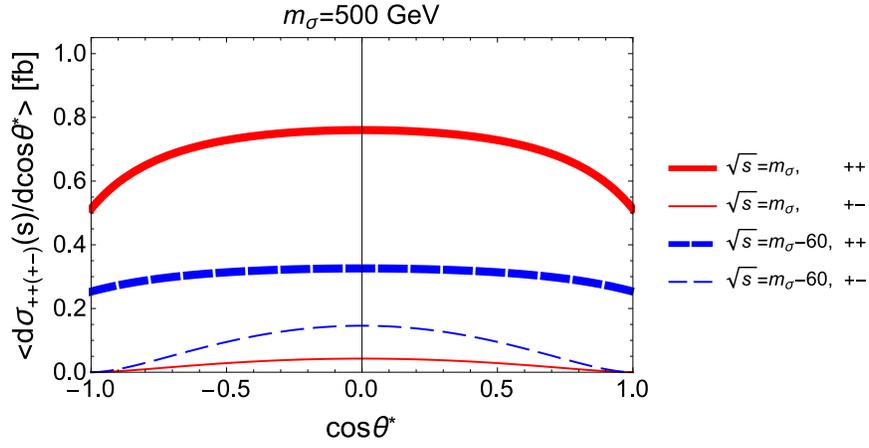}
 \caption{The angular distributions of the luminosity-weighted cross sections
 evaluated at $\sqrt{s_{\gamma\gamma}} =
 m_\sigma$ (thick lines) and $m_\sigma -60$ GeV (thin lines) for
 $m_\sigma = 500$ GeV ($\sqrt{s_{ee}}$ = 625 GeV) case. The solid-red and dashed-blue lines represent
 $(++)$ and $(+-)$ helicity configurations of initial photons,
 respectively.}
 \label{fig:bghh}
\end{figure}

We approximate this expression by neglecting the angular dependence
of the Higgs production cross section. In Fig.~\ref{fig:bghh}, we plot
the luminosity-weighted differential cross section 
\begin{align}
\Big\langle \frac{d\hat{\sigma}_{\lambda \lambda^{'}}\left(\gamma
\gamma\rightarrow hh, s_{\gamma\gamma} ,\theta^* \right)}{d\cos\theta^*} \Big\rangle &\equiv 
\frac{d\hat{\sigma}_{\lambda \lambda^{'}}\left(\gamma \gamma\rightarrow hh; s_{\gamma\gamma} ,\theta^* \right)}{d\cos\theta^*}
\left[ \frac{1}{L_{ee}} \frac{d L_{\gamma \gamma}}{dz} \frac{1 \pm \xi_2 \xi'_2}{2} \right]\!\!\left(z=\sqrt{s_{\gamma\gamma}/s_{ee}}\right),
\end{align}
for $\sqrt{s_{\gamma\gamma}} = m_\sigma$ and $m_\sigma -60\,{\rm GeV}$ with $m_\sigma =
500$ GeV (Point 1).
In evaluating the differential cross section $d\hat{\sigma}_{\lambda
\lambda^{'}}(\gamma\gamma\rightarrow hh; s_{\gamma\gamma} ,\theta^*) / d\cos\theta^*$, we use the
one-loop expressions given by Ref.~\cite{Jikia:1992mt}.

From the figure, we see that the luminosity-weighted
differential cross
sections for the $(++)$ photon helicity are larger than those for
the $(+-)$ photon helicity. Those cross sections do not change significantly over the
whole range of $\cos\theta^*$ for both $\sqrt{s_{\gamma\gamma}}$ choices.
We have also checked that Point 4 shows a similar behavior with $\sqrt{s_{\gamma\gamma}} = 800$ GeV. 
Therefore, we approximate that Higgs pairs are produced almost
isotropically
in the c.m.\ frame of the photon collision. Then,
Eq.~(\ref{eq:sigma_est}) is written as

\begin{align}
\sigma_{\rm cut}(\gamma\gamma\rightarrow hh\rightarrow
 bb\bar{b}\bar{b}; s_{ee}) &\simeq {\rm Br}(h \rightarrow b\bar{b})^2 \nonumber\\
\times \sum_{\lambda
 \lambda^{'} = ++, +-} 
 \int_0^{y_m} dz 
 &\left[ \frac{1}{L_{ee}} \frac{d L_{\gamma \gamma}}{dz}
 \left(
 \frac{1\pm \xi_2 \xi'_2}{2} 
 \right)\, 
 \varepsilon_{\lambda \lambda^{'}} \right]\!\!(z)\,\,
 \frac{d\hat{\sigma}_{\lambda \lambda^{'}}
  (\gamma\gamma\rightarrow hh; s_{\gamma\gamma} = z^2 s_{ee})}{d\cos\theta^*}\Bigg|_{\rm ave.}, 
   \label{eq:approx_hhxsec}
\end{align}
where
\begin{align}
\frac{d\hat{\sigma}_{\lambda \lambda^{'}}
  (\gamma\gamma\rightarrow hh ; s_{\gamma\gamma})}{d\cos\theta^*}\Bigg|_{\rm ave.} &\equiv
 \int^1_{0} d\cos\theta^* \,\frac{d\hat{\sigma}_{\lambda \lambda^{'}}
  (\gamma\gamma\rightarrow hh ; s_{\gamma\gamma} ,\theta^*)}{d\cos\theta^*},\\[+5pt]
 \left[ \frac{1}{L_{ee}} \frac{d L_{\gamma \gamma}}{dz} 
 \left(
 \frac{1\pm \xi_2 \xi'_2}{2} 
 \right)\,
 \varepsilon_{\lambda \lambda^{'}} \right]\!\!(z)\,\,
   &\equiv \int^{y_m}_{z^2/y_{m}} dy \int^1_{0} d\cos\theta^*
   \frac{1}{L_{ee}} \frac{d^2 L_{\gamma \gamma}}{dz dy} \left(\frac{1
 \pm \xi_2 \xi'_2}{2}\right)
\, \varepsilon_{\lambda \lambda^{'}}(z,y,\theta^*).
\end{align}
Note that we approximate the differential cross section by its averaged
value over $\cos\theta^*$.
The total cut efficiency is estimated by generating event samples of 
isotropically produced Higgs pairs
 with the luminosity function, setting the c.m.\ energy of
the Higgs pairs to $\sqrt{s_{\gamma\gamma}}$ 
and imposing all the selection cuts on the generated events.

Finally, we comment on our approximation 
that the Higgs pairs are produced isotropically.
Using the maximum value of 
the differential cross section over the $\cos\theta^*$ range instead of the averaged one in Eq.~\eqref{eq:approx_hhxsec}, 
we obtain the upper bound of $\sigma_{\rm cut}$. 
We check that the differences between the upper bounds and our
approximated cross section, Eq.~\eqref{eq:approx_hhxsec}, 
are less than 15 \%. This can be regarded as the uncertainty of our
approximation,
 which is sufficient for our study
as we will see in the next subsection.

\subsubsection*{$\bm{\gamma \gamma \rightarrow Z Z \rightarrow bb\bar{b}\bar{b}, b\bar{b}c\bar{c}}$}

Next, we discuss the $ZZ$ background. The dominant background
contributions are from $bb\bar{b}\bar{b}$ and $b\bar{b}c\bar{c}$ decay
modes. Instead of directly estimating the $ZZ$ background cross section
with all
the selection cuts being imposed ($\sigma_{\rm cut}$), we set an upper bound
on the cross section by removing the Preselection and S4 cuts since
estimation of the efficiencies of those cuts needs more detailed simulation. 
The upper bound is written as
\begin{align}
 \sigma_{\rm cut}&(\gamma\gamma\rightarrow ZZ\rightarrow
 bb\bar{b}\bar{b}, b\bar{b}c\bar{c}; s_{ee})
\lsim
 \left\{ {\rm Br}(Z\rightarrow b\bar{b})^2 \varepsilon^{4b}_{\rm
 S2} + 
2\,{\rm Br}(Z\rightarrow b\bar{b}) {\rm Br}(Z\rightarrow c\bar{c})
 \varepsilon^{2b2c}_{\rm S2} \right\}\,\varepsilon_{\rm S3} \nonumber\\[+3pt]
&\hspace{1em}\times \sum_{\lambda \lambda^{'} = ++, +-} \int^{\sqrt{s_{\rm max}/s_{ee}}}_{\sqrt{s_{\rm min}/s_{ee}}} dz
\,\left[ \frac{1}{L_{ee}} \frac{d L_{\gamma \gamma}}{dz} \frac{1
 \pm \xi_2 \xi'_2}{2} \right]\!\!\left(z\right)
\hat{\sigma}_{\lambda \lambda^{'}} \left(\gamma\gamma\rightarrow ZZ; s_{\gamma\gamma} =
 z^2 s_{ee} \right),
\label{eq:sigma_zz_est}
\end{align}
where $\varepsilon^{4b(2b2c)}_{\rm S2}$ and $\varepsilon_{\rm S3}$ are the efficiencies of the S2 cut for
the $bb\bar{b}\bar{b}\,(b\bar{b}c\bar{c})$ decay mode and the S3
cut, respectively. In evaluating $\hat{\sigma}_{\lambda \lambda^{'}}
\left(\gamma\gamma\rightarrow ZZ; s_{\gamma\gamma} \right)$, we use the one-loop expressions given in Ref.~\cite{Gounaris:1999hb}.
In
Eq.~\eqref{eq:sigma_zz_est}, we approximately take into account the efficiencies of the S1,
S2 and S3 cuts as follows. The effect of the S1 cut is approximated by limiting the
integration interval, setting the upper and lower limits to those of the
S1 cut
, i.e., $\sqrt{s}_{\rm max} =
m_{\sigma} +40$ GeV and $\sqrt{s}_{\rm min} = m_{\sigma} -60$ GeV for
each sample model points. The
efficiency of the S2 cut corresponds to the probability that three or
four jets are $b$-tagged from the $bb\bar{b}\bar{b}$ ($b\bar{b}c\bar{c}$)
final state
 and
is obtained as $\varepsilon^{4b(2b2c)}_{\rm S2} \sim 0.82
\,(0.12)$.
The S3 cut efficiency is estimated from simulated event samples of
 $Z$ boson pairs produced in the $e^+e^-$ collision, setting the c.m.\ energy of the $Z$ boson pair at
$m_\sigma$ since the peak region of the photon-photon
 luminosity is tuned at around this energy. 
This upper bound on the $ZZ$ background will be used in estimating the
upper bound on the total background in the next subsection.
%%%%%%%%%%%%%%%%%%%%%%%%%%%%%%%%%%%%%%%%%%%%%%

\subsection{Results}
\label{sec:results}
We present expected signal and background event numbers with all the
selection cuts imposed for the sample model points
in Table~\ref{table:results}.
\begin{table}[t]
	\begin{center}
	\begin{tabular}{l|rrrr}
	\hline\hline
	& Point 1 & Point 2 & Point 3 & Point 4
	\\
	\hline
	$m_{\tilde{t}_1}$ [ GeV ] & $250$ & $300$ & $350$ & $400$
	\\
	$\sigma( \gamma \gamma \rightarrow \stoponium \rightarrow hh )$ [fb] & 0.34 & 0.26 & 0.2 & 0.18
	\\
	\hline
	signal 			&$14.2$	&$13.5$	&$12.5$	&$12.4$
	\\
	total background & $\lsim$ 3.9 & $\lsim$ 3.2 & $\lsim$ 2.3 &
			 $\lsim$ 2.3	
	\\
	\hspace{1em} non-resonant $hh$	&$2.2$	&$2.1$	&$1.7$	&$1.4$
	\\
	\hspace{1em} $bb\bar{b}\bar{b}$	&$0.5$	&$0.2$	&$0.0$	&$0.0$
	\\
	\hspace{1em} $c\bar{c}b\bar{b}$	&$0.4$	&$0.5$	&$0.2$	&$0.1$
	\\
	\hspace{1em} $ZZ$			&$\lsim0.8$	&$\lsim0.5$	&$\lsim0.3$	&$\lsim0.7$
	\\
	\hspace{1em} $t\bar{t}$			&$0.1$	&$0.0$	&$0.1$	&$0.1$
	\\
	\hspace{1em} $cc\bar{c}\bar{c}$	&$0.0$	&$0.0$	&$0.0$	&$0.0$
	\\
	\hspace{1em} $b\bar{b}q\bar{q}$	&$0.0$	&$0.0$	&$0.0$	&$0.0$
	\\
	\hspace{1em} $W^+W^-$		&$0.0$	&$0.0$	&$0.0$	&$0.0$
	\\
	\hspace{1em} $W^+W^-Z$		&$0.0$	&$0.0$	&$0.0$	&$0.0$
	\\
	\hline
	Significance $Z_0$	&$\gsim 5.2$	&$\gsim 5.3$	&$\gsim
		     5.5$	&$\gsim 5.5$
	\\
	\hline\hline
	\end{tabular}
	\caption{The number of the signal and background events that pass all the selection cuts 
	and the signal significance $Z_0$ with $\mathcal{L}_{ee} = 1
	 \rm{ab}^{-1}$ for each sample point. }
	\label{table:results}
	\end{center}
\end{table}
Here, we assume the integrated electron-beam luminosity of $1\mspace{5mu}\textrm{ab}^{-1}$.  
More than ten signal events are expected for all the sample points,
while background events are effectively reduced to less than four events.
We estimate the expected significance of detecting the $\stoponium
\rightarrow hh$ decay mode using an approximated formula based on the Poisson
distribution~\cite{Cowan:2010js}:
\begin{align}
Z_0 = \sqrt{2\left\{ (S+B) \ln(1+S/B) -S \right\}},
\end{align}
with $S$ $(B)$ being the expected signal (total background) event
number.\footnote{This significance approaches to $S/\sqrt{B}$ when $S \ll B$.}
The significance $Z_0$ for each sample point is also presented in Table~\ref{table:results}.
Because we only estimate
the upper bounds on the $ZZ$ background, the expected
significances are regarded as lower bounds.
We see 
that in order for the $5\sigma$ detection, the signal cross sections, 
$\sigma( \gamma \gamma \rightarrow \stoponium \rightarrow hh )$, of $ 0.34
- 0.18$ fb are required for the stoponium masses of 500 --
800 GeV, respectively. 
In the rest of this section, we discuss how background events are
reduced by imposing the selection cuts.

After imposing all the selection cuts, 
the major background source is the non-resonant Higgs pair ($hh$)
production process, and the contributions from
other background sources except $bb\bar{b}\bar{b}$, $b\bar{b}c\bar{c}$
and $ZZ$ are negligibly small.
We present the cut-flow information along with the cut efficiencies in
parentheses
for the sample Point 1 in Table~\ref{table:cuthistory_1} and 
for other points in Table~\ref{table:cuthistory}, where only the non-negligible background processes are presented.
\begin{table}[t]
	\begin{center}
	\begin{tabular}{l|rrrr}
	\hline\hline
	Point 1 & Preselection $+$ S1 & $+$ S2 & $+$ S3 & $+$ S4 
	\\
	\hline
	signal 			&$32.8$	&$26.8$ {\footnotesize$(0.82)$}	&$18.0$ {\footnotesize$(0.67)$}	&$14.2$ {\footnotesize$(0.79)$}
	\\
	non-resonant $hh$	&$4.9$ 	&$4.0$ {\footnotesize$(0.82)$}	&$2.8$ {\footnotesize$(0.70)$}	&$2.2$ {\footnotesize$(0.76)$}
	\\
	$bb\bar{b}\bar{b}$	&$42$	&$34$ {\footnotesize$(0.82)$}	&$1.9$ {\footnotesize$(0.05)$}	&$0.5$ {\footnotesize$(0.25)$}
	\\
	$c\bar{c}b\bar{b}$	&$540$ 	&$67$ {\footnotesize$(0.12)$}	&$2.1$ {\footnotesize$(0.03)$}	&$0.4$ {\footnotesize$(0.17)$}
	\\
	$ZZ$ 	& - & - & - &$\lsim0.8$
	\\
	$t\bar{t}$			&$39$	&$2.2$ {\footnotesize$(0.06)$}	&$0.1$ {\footnotesize$(0.04)$}	&$0.1$ {\footnotesize$(1.00)$}
	\\
	$cc\bar{c}\bar{c}$	&$1160$	&$4.3$ {\footnotesize$(0.004)$}	&$0.2$ {\footnotesize$(0.04)$}	&$0.0$ {\footnotesize$(0.17)$}
	\\
	$b\bar{b}q\bar{q}$	&$1190$	&$1.5$ {\footnotesize$(0.001)$}	&$0.1$ {\footnotesize$(0.04)$}	&$0.0$ {\footnotesize$(0.25)$}
	\\
	$W^+W^-$		&$195800$	&$3.9$ {\footnotesize$(2\times10^{-5})$}	&$0.0$ {\footnotesize$(0.00)$}	&$0.0$ {\footnotesize$(0.00)$}
	\\
	$W^+W^-Z$		&$5.0$	&$0.2$ {\footnotesize$(0.04)$}	&$0.0$ {\footnotesize$(0.02)$}	&$0.0$ {\footnotesize$(1.00)$}
	\\
	\hline\hline
	\end{tabular}
	\caption{The number of the signal and background events 
	after the successive application of the cuts 
	with $\mathcal{L}_{ee} = 1 \rm{ab}^{-1}$ for the sample Point 1. The efficiencies of
	 each selection cuts are also presented in the parentheses. 
	For the $ZZ$ background, only an upper bound on the number of events that pass all the cuts is presented.} 
	\label{table:cuthistory_1}
	\end{center}
\end{table}
\begin{table}[t]
	\begin{center}
	\begin{tabular}{l|rrrr}
	\hline\hline
	Point 2 & Preselection $+$ S1 & $+$ S2 & $+$ S3 & $+$ S4 
	\\ 
	\hline
	signal 			&$28.6$	&$23.4$ {\footnotesize$(0.82)$}	&$14.6$ {\footnotesize$(0.62)$}	&$13.5$ {\footnotesize$(0.92)$}
	\\
	non-resonant $hh$	&$4.1$ 	&$3.4$ {\footnotesize$(0.82)$}	&$2.3$ {\footnotesize$(0.67)$}	&$2.1$ {\footnotesize$(0.93)$}
	\\
	$bb\bar{b}\bar{b}$	&$30$ 	&$25$ {\footnotesize$(0.82)$}	&$0.3$ {\footnotesize$(0.01)$}	&$0.2$ {\footnotesize$(0.54)$}
	\\
	$c\bar{c}b\bar{b}$	&$390$ 		&$49$ {\footnotesize$(0.12)$}	&$1.1$ {\footnotesize$(0.02)$}	&$0.5$ {\footnotesize$(0.43)$}
	\\
	\hline\hline
	Point 3 & Preselection $+$ S1 & $+$ S2 & $+$ S3 & $+$ S4 
	\\
	\hline
	signal 			&$23.2$ 	&$19.0$ {\footnotesize$(0.82)$}	&$12.9$ {\footnotesize$(0.68)$}	&$12.5$ {\footnotesize$(0.97)$}
	\\
	non-resonant $hh$	&$3.1$ 	&$2.5$ {\footnotesize$(0.82)$}	&$1.8$ {\footnotesize$(0.70)$}	&$1.7$ {\footnotesize$(0.97)$}
	\\
	$bb\bar{b}\bar{b}$	&$24$ 	&$19$ {\footnotesize$(0.82)$}	&$0.1$ {\footnotesize$(0.01)$}	&$0.0$ {\footnotesize$(0.00)$}
	\\
	$c\bar{c}b\bar{b}$	&$300$ 	&$37$ {\footnotesize$(0.12)$}	&$0.3$ {\footnotesize$(0.01)$}	&$0.2$ {\footnotesize$(0.50)$}
	\\
	\hline\hline
	Point 4 & Preselection  $+$ S1 & $+$ S2 & $+$ S3 & $+$ S4 
	\\
	\hline
	signal 			&$21.6$ 	&$17.7$ {\footnotesize$(0.82)$}	&$12.6$ {\footnotesize$(0.71)$}	&$12.4$ {\footnotesize$(0.99)$}
	\\
	non-resonant $hh$	&$2.4$ 	&$2.0$ {\footnotesize$(0.82)$}	&$1.4$ {\footnotesize$(0.71)$}	&$1.4$ {\footnotesize$(0.99)$}
	\\
	$bb\bar{b}\bar{b}$	&$17$  	&$14$ {\footnotesize$(0.82)$}	&$0.1$ {\footnotesize$(0.01)$}	&$0.0$ {\footnotesize$(0.00)$}
	\\
	$c\bar{c}b\bar{b}$	&$230$ 	&$28$ {\footnotesize$(0.12)$}	&$0.3$ {\footnotesize$(0.01)$}	&$0.1$ {\footnotesize$(0.50)$}
	\\
	\hline\hline
	\end{tabular}
	 \caption{Same as Table~\ref{table:cuthistory_1} but for the
	 sample Point 2, 3 and 4, and only the non-negligible background processes
	 are shown.}
	\label{table:cuthistory}
	\end{center}
\end{table}

The selection cut S2, 
which requires three or four jets are b-tagged, then plays an important role 
in reducing the large portion of the background events which needs some
non-$b$ jets to be mistagged to pass the cut.

The selection cut S3, relevant to the di-jet invariant masses, also
reduce most of the background events 
efficiently, except for non-resonant $hh$, by imposing Higgs mass
constraints on two pairs of jets.
At this stage, only the non-resonant $hh$, $bb\bar{b}\bar{b}$,
$c\bar{c}b\bar{b}$ and $ZZ$ backgrounds remains sizable.

\begin{figure}[t]
 \centering
 \begin{subfigure}{0.45\columnwidth}
  \centering
  \includegraphics[width=\columnwidth]{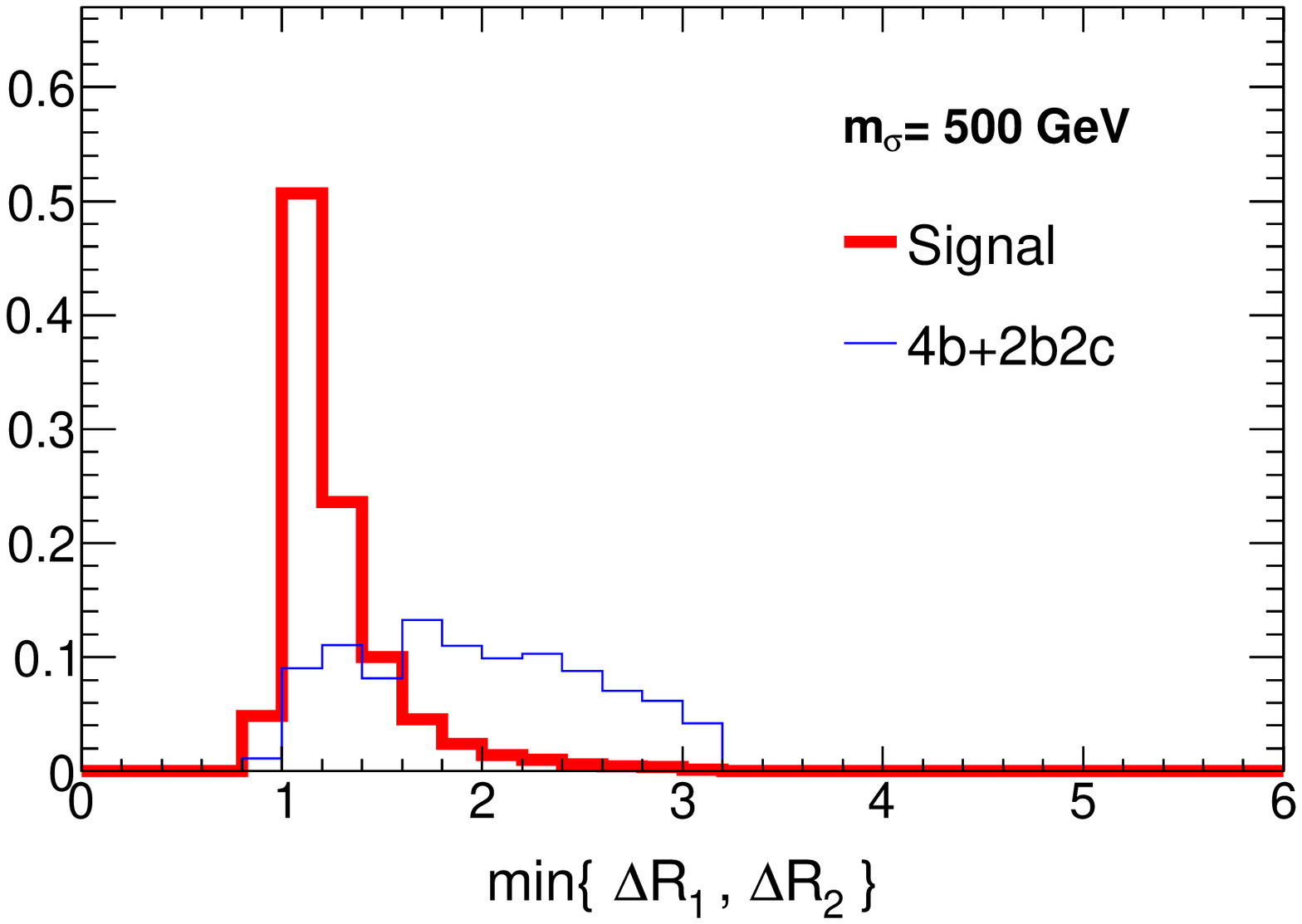}
  \caption{$\min\{ \Delta R_1 , \Delta R_2 \}$}
 \end{subfigure}
 \begin{subfigure}{0.45\columnwidth}
  \centering
  \includegraphics[width=\columnwidth]{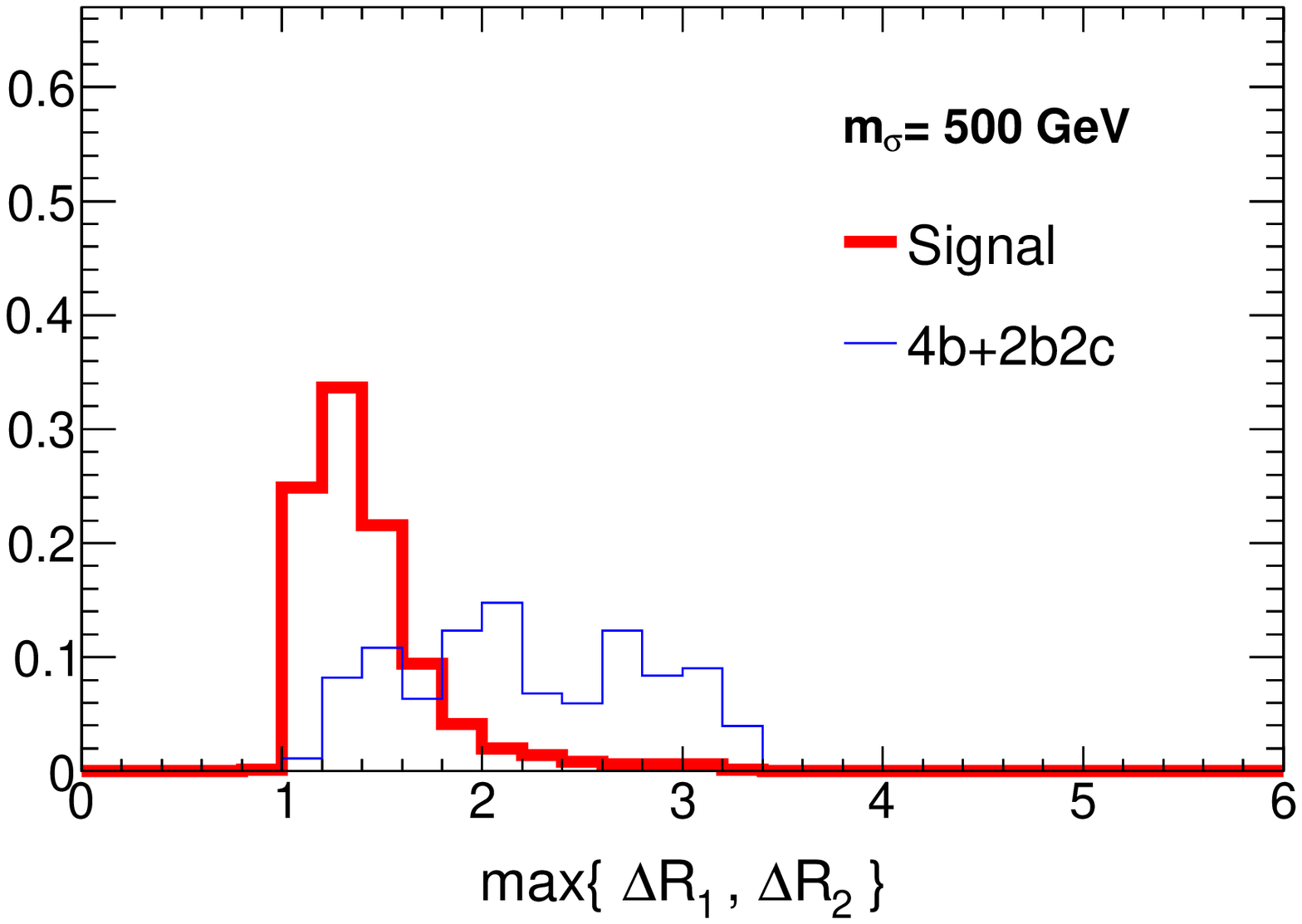}
  \caption{$\max\{ \Delta R_1 , \Delta R_2 \}$}
 \end{subfigure}
 \caption{The normalized distributions of $\min\{ \Delta R_1 , \Delta R_2 \}$ (left) and 
 	$\max\{ \Delta R_1 , \Delta R_2 \}$ (right) for the Point 1 with
  all the cuts except S4 being imposed.
	The thick red lines are for the signal and 
	the thin blue lines for the $bb\bar{b}\bar{b}$ plus $c\bar{c}b\bar{b}$.}
\label{fig:dr1}
\end{figure}

The selection cut S4, which is based on $\Delta R$ distributions, further reduces 
the remained $bb\bar{b}\bar{b}$ and $c\bar{c}b\bar{b}$ backgrounds.
In Figs.~\ref{fig:dr1}, 
we present the $\min \{ \Delta R_1, \Delta R_2 \}$ and $\max \{ \Delta R_1, \Delta R_2 \}$ distributions of 
signal and $bb\bar{b}\bar{b}$ plus $c\bar{c}b\bar{b}$ background 
after imposing the Preselection, S1, S2 and S3 cuts for the sample Point 1.
We see that $\Delta R$ tends to be small for
the signal, 
while it can be large up to
$\sim 3$ 
for the $bb\bar{b}\bar{b}$ plus $b\bar{b}c\bar{c}$ backgrounds.
This difference makes the S4 cut efficient for reducing those backgrounds,
and can be understood qualitatively as follows.
The di-jet systems from the non-resonant multi-jet processes tend to distribute
 in the large $|\eta|$ region more than the di-jet systems from decays of rather
isotropically produced Higgs bosons.
 In general,
two jets in a di-jet system with larger $|\eta|$ tend to have larger azimuthal angle
difference, $\Delta \phi$, and thus larger $\Delta
R$; this mainly makes the difference in the $\Delta R$ distributions between the
signal and the four-jet backgrounds above.

%%%%%%%%%%%%%%%%%% 
\section{Implication to the stop sector}
In the previous section, we have shown that there are possibilities to
detect the di-Higgs decay mode of the stoponium and measure its
cross section, $\sigma
    (\gamma\gamma\rightarrow \stoponium \rightarrow hh)$.
In this section we discuss its implication for
extracting information on
 the SUSY parameters in the stop sector: $m_{\tilde{t}_2}$ and $A_t$.
Since we assume that the lighter stop and the lightest neutralino are
discovered by the time when the photon-photon collider experiment 
will be carried out, their masses are regarded as known parameters. Some other
SUSY parameters may also be known by that time, but we just assume them
as unknown parameters for a conservative approach.  

\begin{figure}[t]
 \centering
  \includegraphics[width=0.4\columnwidth]{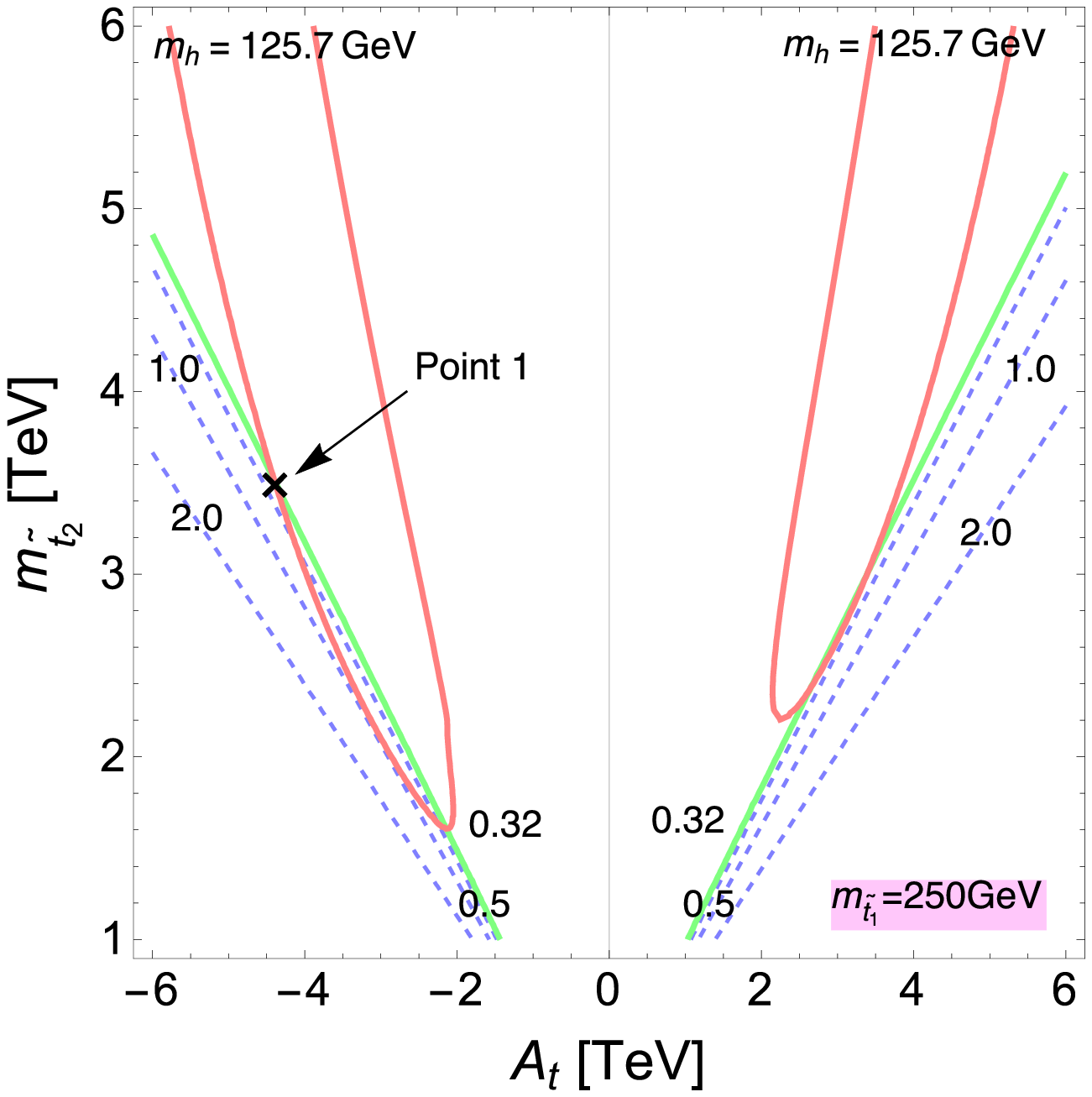}
  \includegraphics[width=0.4\columnwidth]{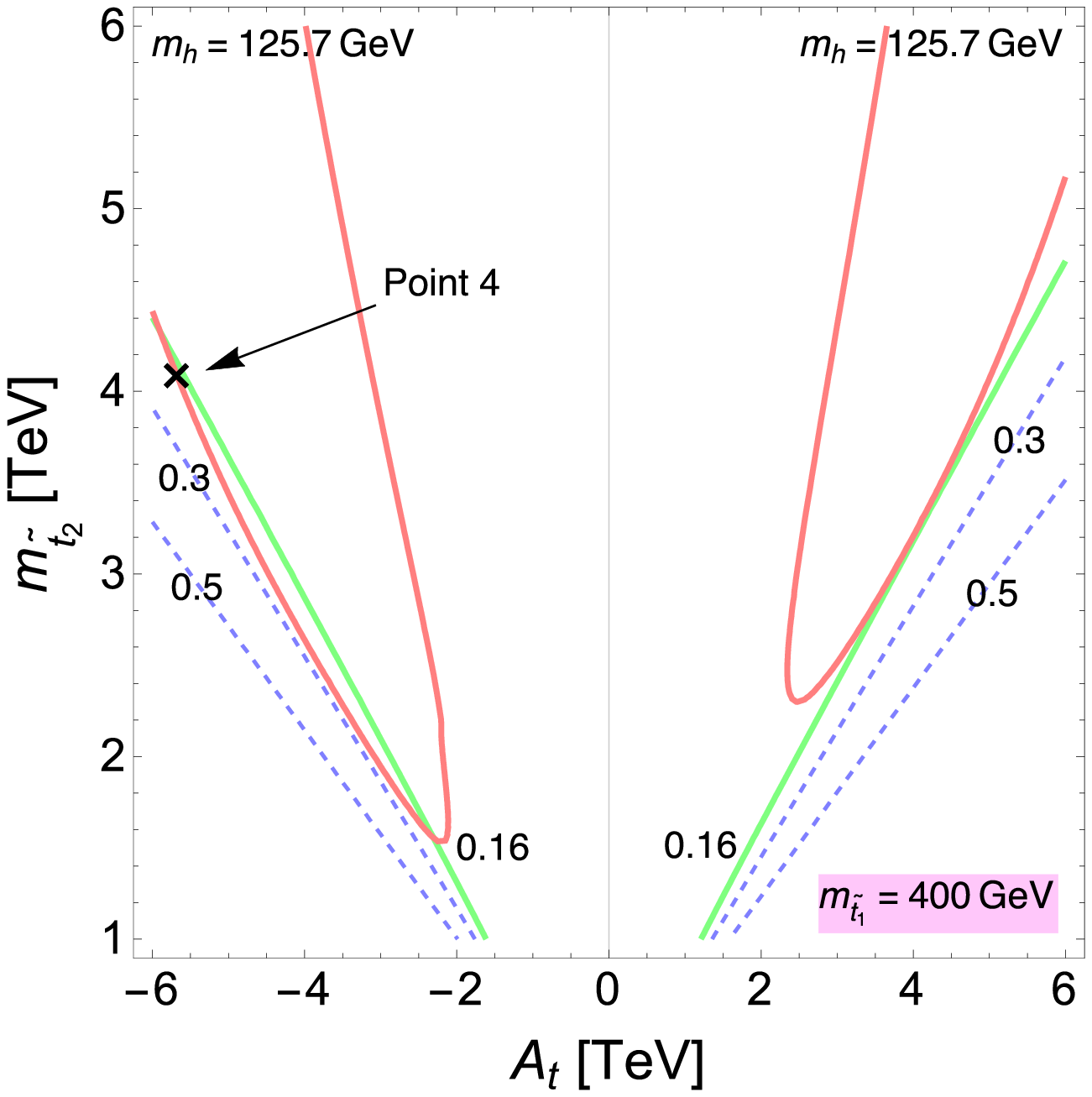}
   \caption{\small The contours of the stoponium cross section,
    $\sigma(\gamma\gamma\rightarrow\stoponium\rightarrow hh)$, and $m_h =
 125.7$ GeV on the $m_{\tilde{t}_2}$--$A_t$ plane for $m_{\tilde{t}_1}=250\ {\rm GeV}$
 (left) and 400 GeV (right). We take $m_{\tilde{\chi}^0_1}=150$ GeV (left) and 350 GeV (right), $\tan\beta=10$, $\mu=2\ {\rm TeV}$ 
    and other SUSY parameters having dimension of mass equal to 2 TeV,
 except the trilinear couplings of the first and second generation
 sfermions which are set to zero.
    The dashed-blue lines show the cross sections, while
 the solid-green line represents the cross section which
 allows $5 \sigma$ detection of the $\stoponium \rightarrow
 hh$ mode. The contour labels are given in units of fb. The red curve
 shows the Higgs boson mass constraint. The sample model points in
 Sec.~\ref{section:collider_study} are shown with the cross symbol.} 
  \label{fig:cs}
\end{figure}

The heavier stop mass, $m_{\tilde{t}_2}$, and
stop trilinear coupling, $A_t$, may be determined from the constraints of the
measured cross section and Higgs mass up to the four-fold solutions when we fix the other SUSY
parameters. 
This is illustrated in Fig.~\ref{fig:cs}, where the
four solutions appear on the $m_{\tilde{t}_2}$--$A_t$ plane as the
intersections between the contours of the stoponium cross section
and Higgs mass.

Since we assume that the true SUSY parameter values are not known (except
$m_{\tilde{t}_1}$ and $m_{\chi^0_1}$), we scan over the unknown SUSY
parameters relevant to the stoponium cross sections and Higgs mass,
finding possible solutions of $m_{\tilde{t}_2}$ and $A_t$ in the parameter space.
As an example result, we discuss upper bounds on $m_{\tilde{t}_2}$ and $|A_t|$.\footnote{Lower
bounds could also be derived in the same way; however, they
depend on the uncertainty of the Higgs mass significantly, and we do not
consider them in the following discussion. 
}
For the paramter scan, we employ the phenomenological MSSM~\cite{Berger:2008cq} as a SUSY framework.
Besides $\tan\beta$ and $\mu$ parameters, which are relevant to the
stoponium cross section and Higgs mass at tree level, the
parameter space is spanned by the mass parameters of sbottom, stau and gluino:
 $M_{\tilde{b}_R}, M_{\tilde{\tau}_L}, M_{\tilde{\tau}_R}$ and $M_3$.
For simplicity, we set $M_{\tilde{\tau}_L} = M_{\tilde{\tau}_R}$ 
and fix all the other SUSY mass parameters and trilinear couplings 
to 2 TeV.
We check that the bounds are insensitive to these assumptions.
In Table~\ref{table:parameters}, we summarize the scanned and assumed
SUSY parameters for each $m_{\tilde{t}_1}$. 
\begin{table}[t]
  \begin{center}
    \begin{tabular}{c|c|c|c|c}
      \hline\hline
      $m_{\tilde{t}_1}$ [GeV] & 250 & 300 & 350 & 400
      \\
      $m_{\tilde{\chi}^0_1}$ [GeV] & 150 & 250 & 300 & 350
      \\
      $\sigma( \gamma \gamma \rightarrow \stoponium \rightarrow hh )$ [fb] & 0.34 & 0.26 & 0.2 & 0.18
      \\
      \hline
      $\tan\beta$ & \multicolumn{4}{|c}{[10, 60] (3 points)}
      \\
      $\mu$ & \multicolumn{4}{|c}{[-10, 10] TeV (6 points)}
      \\
      $M_3$ & \multicolumn{4}{|c}{[2, 10] TeV (3 points)}
      \\
      $M_{\tilde{b}_R}$ & \multicolumn{4}{|c}{[2, 10] TeV (3 points)}
      \\
      $M_{\tilde{\tau}_L} = M_{\tilde{\tau}_R}$ & \multicolumn{4}{|c}{[2, 10] TeV (3 points)}
      \\
      $M_{{\rm others}}$ & \multicolumn{4}{|c}{2 TeV}
      \\
      $A_b = A_{\tau}$ & \multicolumn{4}{|c}{2 TeV}
      \\
      $A_{{\rm 1st}/{\rm 2nd}}$ & \multicolumn{4}{|c}{2 TeV}
      \\
      \hline\hline
    \end{tabular}
    \caption{The assumed SUSY parameter values and the ranges of the scanned
   parameters. The parentheses show the number of points
   taken for the corresponding parameter. The stoponium cross sections are set such that they allow $5
   \sigma$ detection of the $\gamma \gamma \rightarrow \stoponium
   \rightarrow hh$ process. $M_{\rm others}$ represents the SUSY mass
   parameters which are not explicitly shown in the Table. $A_{{\rm
   1st}/{\rm 2nd}}$ denotes the trilinear couplings
   related to the first and second generation
   sfermions.}
    \label{table:parameters}
  \end{center}
\end{table}
We assume that the $\gamma \gamma \rightarrow \stoponium \rightarrow hh$
process will be measured with more than $5 \sigma$ significance and regard
the cross sections given in Table~\ref{table:results} as the measured
ones for each $m_{\tilde{t}_1}$.
In obtaining bounds, we take into account statistical uncertainties
of the signal measurements. 

As shown in Table~\ref{table:upper_bound}, the obtained upper bounds on $m_{\tilde{t}_2}$ and $|A_t|$ are $3.8
- 4.7$ TeV and $5.1 - 6.5$ TeV, respectively.\footnote{The upper bounds on $|A_t|$
are from the negative-large solutions, while the positive-large solutions
also exist up to the similar, but $\sim 1$ TeV narrower, $|A_t|$ range.} 
\begin{table}[t]
  \begin{center}
    \begin{tabular}{l|rrrr}
      \hline\hline
      $m_{\tilde{t}_1}$ [GeV] & 250 & 300 & 350 & 400
      \\
      \hline
      $m_{\tilde{t}_2}^{\rm upper}$ [TeV] & 3.8 & 4.2 & 4.6 & 4.7
      \\
      $|A_t|^{\rm upper}$ [TeV] & 5.1 & 5.7 & 6.2 & 6.5
      \\
      \hline\hline
    \end{tabular}
    \caption{
    The obtained upper bounds on $m_{\tilde{t}_2}$ and $|A_t|$ by the
   parameter scan for each
   $m_{\tilde{t}_1}$ value.
    }
    \label{table:upper_bound}
  \end{center}
\end{table}
The upper bounds on $m_{\tilde{t}_2}$ are obtained well within the
scanned parameter space and do not change significantly even if we extend
the parameter space to $|\mu|, M_3, M_{\tilde{b}_R}, M_{\tilde{\tau}} <
14$ TeV from 10 TeV. 
On the other hand, the upper bounds on $|A_t|$
increase non-negligibly as we extend the parameter space; the dominant
effects on the $|A_t|$ bounds
are from the change of the $\mu$
parameter range since the $A_t$ parameter
linearly depends on the $\mu$ parameter through $A_t = X_t -\mu
\cot\beta$.% Eq.~\eqref{eq:Xt}. 
\,Thus, information on the $\mu$ parameter is important to obtain
stringent bounds on $A_t$.
As illustrated in this section, detection of the di-Higgs decay mode of the stoponium and
measurement of its cross section would provide useful information on the stop sector.
%%%%%%%%%%%%%%%%%%%%%%%%%%%%%%%%%%%%%%%%%%%%%%%%

\section{Summary and conclusion}
\label{conclusion}
In this study, we have investigated the detectability of the stoponium
in the di-Higgs decay mode at the photon-photon collider. We have
assumed that the lightest neutralino is the LSP, and the lighter stop is
the next-to-lightest SUSY particle (NLSP). We have also assumed that
those particles would be discovered before the photon-photon collider experiment will be carried
out and that
the basic properties of the stop such as the mass and left-right
mixing angle could
be studied by that time. We have concentrated on the scenario where the mass
difference between the stop and the neutralino is small
enough, and the stop can form the stoponium. 

The detectability of the stoponium di-Higgs decays has been investigated
by estimating the 
stoponium signal and standard model backgrounds and optimizing the signal selection cuts.
It has been found that $5\sigma$ detection of the di-Higgs decay mode
is possible with the integrated
electron-beam luminosity of 1 ${\rm ab}^{-1}$ if the signal cross
section, $\sigma(\gamma \gamma \ra \stoponium \ra hh)$, of 0.34, 0.26,
0.2 and 0.18
fb are realized for the stoponium masses of 500, 600, 700 and 800 GeV, respectively. As concrete
examples, we have
provided the four sample model points in MSSM, corresponding to those
stoponium masses and realizing such cross sections. 
 
Finally, we have discussed the implication of the cross section
measurement of the stoponium di-Higgs decay mode for the MSSM stop sector. Combining the measured cross section
with the Higgs-mass constraint, we have shown that there would be the
upper bound on the heavier stop mass for each lighter stop and lightest
neutralino masses. $A_t$ parameter would also be constrained, depending on 
other SUSY parameters such as $\mu$ and $\tan\beta$.      
In conclusion, there are possibilities that the di-Higgs decay mode of
the stoponium would be observed unambiguously at the future photon-photon collider and
provide new insights into the stop sector.

\vspace{1em}
\noindent {\it Acknowledgments}: The work is supported by Grant-in-Aid
for Scientific research Nos.\ 23104008 and 26400239.

%\clearpage
\appendix
\section{Photon luminosity function}
\label{app:luminosityfn}
We use the luminosity function of the following
form~\cite{Ginzburg:1981vm,Ginzburg:1982yr,Ginzburg:1999wz}:
\begin{align}
  \frac{1}{L_{ee}} \frac{d^2L_{\gamma\gamma}}{dydy'} = 
  f(x, y) B(x, y) f(x, y') B(x, y').
\end{align}
The function $f$ is given by
\begin{align}
  f(x, y) = \frac{2\pi\alpha_e^2}{\sigma_c x m_e^2} C_{00}(x,y),
\end{align}
where the function $C_{00}$ is given in Eq.\ \eqref{C00}, and
\begin{align}
  \sigma_c = \sigma_c^{\rm (np)} + \lambda_e P_l \sigma_1,
\end{align}
with 
\begin{align}
  \sigma_c^{\rm (np)} &= \frac{2\pi \alpha_e^2}{x m^2_e} \left[ \left( 1 - \frac{4}{x} - \frac{8}{x^2} \right) \ln(x+1) + \frac{1}{2} + \frac{8}{x} - \frac{1}{2(x+1)^2}\right],
  \\[+3pt]
  \sigma_1 &= \frac{2\pi \alpha_e^2}{x m^2_e} \left[ \left( 1 + \frac{2}{x} \right) \ln(x+1) - \frac{5}{2} + \frac{1}{x+1} - \frac{1}{2(x+1)^2}\right]. 
\end{align}
The function $B$ is given by
\begin{align}
  B (x,y) = 
  \left\{ 
    \begin{array}{ll}
      \displaystyle{
        \exp 
        \left[ -\frac{\rho^2}{8} \left( \frac{x}{y} - x - 1 \right) \right]
      }
      & ~:~ y_m/2 < y < y_m
      \\
      0 & ~:~ \mbox{otherwise}
    \end{array}
  \right. ,
\end{align}
with $y_m=x/(x+1)$.  In our numerical calculation, we take $\rho=1$
\cite{Ginzburg:1999wz}.

\section{Matrix elements}
\label{app:melements}
We summarize the matrix elements for the stop anti-stop annihilation
processes used in our study~\cite{Drees:1993uw, Martin:2008sv}.
We assume that all the SUSY particles, except the stops, sbottoms and lightest neutralino, are sufficiently heavy, and 
neglect their contributions.
We also assume that the lightest neutralino is purely bino-like.
%Although omitted in the following, 
In the following expressions, the summations over the color indices of
 the initial stop and anti-stop have been implicitly performed as
\begin{equation}
%\left| \mathcal{M} \left( \tilde{t}_1 \tilde{t}^*_1 \rightarrow AB
 %\right) \right|^2_{v=0}
%=
%\left| \frac{1}{3} \sum_{a} \mathcal{M}^{\lambda_A, \lambda_B} 
%\left( \tilde{t}_1 ( \bm{k}=\bm{0} , a )\, \tilde{t}^*_1 ( -\bm{k}=\bm{0}, a ) 
%\rightarrow A B \right) \right|^2 .
\left| \mathcal{M} \left( \tilde{t}_1 \tilde{t}^*_1 \rightarrow AB \right) \right|^2
=
\left| \frac{1}{3}\sum_{a} \mathcal{M}
\left( \tilde{t}_1^a\, \tilde{t}_1^{a*}
\rightarrow A B \right) \right|^2,
\end{equation}
and the explicit color summations should be taken for the final-state
colored paricles.

\subsubsection*{(1) $\bm{gg}$}

In the $v=0$ limit (where $v$ is the velocity of the stops in the
initial state), the contributions from the $t$- and $u$-channel stop exchanges are absent.
Therefore, the squared matrix element dose not depend on the MSSM parameters and is given by
\begin{align}
	\sum_{\rm{color, \,helicity}} \left| \mathcal{M} 
	\left( \tilde{t}_1 \tilde{t}^*_1 \rightarrow gg \right) \right|^2_{v=0} = 
	\left(\frac{16 \pi}{3} \alpha_s \right)^2.
\end{align}

\subsubsection*{(2)  $\bm{\gamma \gamma}$}

As in the $gg$ final-state case, the squared matrix element
for the $\gamma\gamma$ final state dose not depend on the MSSM parameters and is given by
\begin{equation}
	\sum_{\rm{helicity}} \left| \mathcal{M}
	 \left( \tilde{t}_1 \tilde{t}^*_1 \rightarrow \gamma \gamma \right) \right|^2_{v=0} = 
	 128 \pi^2 \left(\frac{2}{3}  \right)^4 \alpha^2_e.
\end{equation}

\subsubsection*{(3) $\bm{ hh }$}

The squared matrix element is given by
\begin{align}
\left| \mathcal{M} \left( \tilde{t}_1 \tilde{t}^*_1 \rightarrow hh \right) \right|^2_{v=0} 
= \Biggl\{
 \frac{2 \left( c^{(2)}_{\tilde{t}_1} \right)^2 }{ 2 m^2_{\tilde{t}_1} - m^2_h} &+
	 \frac{ 2\left( c^{(2)}_{\tilde{t}_1 \tilde{t}_2} \right)^2 }{ m^2_{\tilde{t}_1}+m^2_{\tilde{t}_2}-m^2_h }+
	 c^{22}_{11} \notag \\[+3pt]
	 &+\frac{ c^{(2)}_{\tilde{t}_1} }{ 4 m^2_{\tilde{t}_1} - m^2_h } 
	 \frac{ 3 g \,m_Z }{ 2 c_W } \cos 2\alpha \sin \left( \beta + \alpha \right)
	 \Biggr\}^2 ,
\label{msq_hh}
\end{align}
where $g$ is the $SU(2)_L$ gauge coupling constant, $c_{\rm W} = \cos\theta_{\rm W}$, $s_{\rm W} = \sin\theta_{\rm W}$, and
$\alpha$ is the mixing angle of the CP-even Higgs bosons. In
addition, $c^{(2)}_{\tilde{t}_1}$, $c^{(2)}_{\tilde{t}_1\tilde{t}_2}$ and $c^{22}_{11}$ 
are the coefficients of the
$\tilde{t}_1 \tilde{t}_1^* h$, \,$\tilde{t}_1 \tilde{t}_2^* h$, and 
$\tilde{t}_1 \tilde{t}_1^* h h$ vertices, respectively, and are given by
\begin{align}
	c^{(2)}_{\tilde{t}_1} &= 
	\frac{g \,m_Z}{c_W} \sin \left( \alpha + \beta \right) 
	\left( \frac{1}{2} \cos^2 \theta_{\tilde{t}} -\frac{2}{3} s^2_W \cos 2\theta_{\tilde{t}} \right)
	- \frac{g \,m^2_t}{m_W} \frac{\cos \alpha}{\sin \beta} \notag \\[+3pt]
	&\mspace{294mu}+ \frac{g \,m_t}{2 m_W \sin \beta} \sin 2\theta_{\tilde{t}} 
	\left( A_t \cos \alpha - \mu \sin \alpha \right) ,\\[+8pt]
	c^{(2)}_{\tilde{t}_1\tilde{t}_2} &= 
	\frac{g \,m_Z}{c_W}\sin(\alpha+\beta) \sin 2\theta_{\tilde{t}} 
	\left( \frac{2}{3} s_W^2 -\frac{1}{4} \right)
	+ \frac{g \,m_t}{2 m_W \sin \beta}
	\cos 2\theta_{\tilde{t}} \left( A_t \cos \alpha - \mu \sin \alpha \right) ,\\[+8pt]
	c^{22}_{11}\,\, &= 
	\frac{g^2}{2} \left[ 
	\frac{\cos 2\alpha}{c_W^2} 
	\left( \frac{1}{2} \cos^2 \theta_{\tilde{t}} - \frac{2}{3} s^2_W \cos 2\theta_{\tilde{t}} \right)
	- \frac{m_t^2}{m_W^2} \frac{\cos^2 \alpha}{\sin^2 \beta}
	\right] .
\end{align}

\subsubsection*{(4)  $\bm{ W^+W^- }$}

The squared matrix element is given by
\begin{equation}
	\sum_{\rm{spin}} \left| \mathcal{M} \left( \tilde{t}_1 \tilde{t}^*_1 \rightarrow W^+W^- \right) \right|^2_{v=0} = 
	2 \left( a^T_{WW} \right)^2 + \left( a^L_{WW} \right)^2 ,
\end{equation}	
where $a^T_{WW}$ and $a^L_{WW}$ correspond to the transverse and longitudinal components of the matrix element and are given by
\begin{align}
a^T_{WW} = \mathcal{M}^{+1  +1} =\mathcal{M}^{-1  -1} =
	- \left( \frac{g^2}{2} \cos^2 \theta_{\tilde{t}} - \frac{g_{hWW}  c^{(2)}_{\tilde{t}_1}}{4 m^2_{\tilde{t}_1} - m^2_h} \right),
\end{align}
and
\begin{align}	
a^L_{WW} =\mathcal{M}^{0 0} =
	\left( \frac{2 m^2_{\tilde{t}_1}}{m_W^2} - 1 \right) &
	\left( \frac{g^2}{2} \cos^2 \theta_{\tilde{t}} - 
	\frac{g_{hWW}  c^{(2)}_{\tilde{t}_1}}{4 m^2_{\tilde{t}_1} - m^2_h} \right) \notag \\[+3pt]
	& -2 \left( \frac{m^2_{\tilde{t}_1}}{m_W^2} - 1 \right) \left( 
	\frac{ g^2 \cos^2 \theta_{\tilde{t}} \,m_{\tilde{t}_1}^2 }{ m_{\tilde{t}_1}^2 + m_{\tilde{b}_L}^2 -m_W^2 } \right),
\end{align}
respectively. Here, $m_{\tilde{b}_L}$ is the left-handed sbottom
mass (where we neglect the
left-right sbottom mixing), and $g_{hWW}$ is the coefficient of the $hW^+W^-$ vertex, which is given by
\begin{equation}
g_{hWW} = g \,m_W \sin \left( \beta - \alpha \right) .
\end{equation}
Note that in the $v\rightarrow 0$ limit, the contribution of the $s$-channel $Z$ boson exchange is absent.

\subsubsection*{(5)  $\bm{ ZZ }$}

The squared matrix element is given by
\begin{equation}
	\sum_{\rm{spin}} \left| \mathcal{M}
	\left( \tilde{t}_1 \tilde{t}^*_1 \rightarrow ZZ \right) \right|^2_{v=0} = 
	2 \left( a^T_{ZZ} \right)^2 + \left( a^L_{ZZ} \right)^2 ,
\end{equation}
where
\begin{equation}
a^T_{ZZ} = \mathcal{M}^{+1  +1} =\mathcal{M}^{-1  -1} =
	\frac{1}{c_W^2} \left[ 2g^2 \left\{ \left(\frac{1}{4}-\frac{2}{3}s_W^2 \right) \cos^2 \theta_{\tilde{t}} +\frac{4}{9} s^4_W \right\} 
		-\frac{g_{hWW} c^{(2)}_{\tilde{t}_1}}{ 4 m^2_{\tilde{t}_1} - m^2_h } \right],
\end{equation}
and
\begin{align}
a^L_{ZZ} =\mathcal{M}^{0 0} =
	-\frac{1}{c_W^2} \left( \frac{2 m^2_{\tilde{t}_1}}{m_Z^2} - 1 \right) 
	\left[ 2g^2 \left\{ \left( \frac{1}{4} - \frac{2}{3} s_W^2 \right) \cos^2 \theta_{\tilde{t}} + \frac{4}{9} s_W^4 \right\} 
	- \frac{g_{hWW} c^{(2)}_{\tilde{t}_1}}{ 4 m^2_{\tilde{t}_1} - m^2_h } \right]& \notag\\[+5pt]
	\mspace{20mu}+ \frac{2g^2 m^2_{\tilde{t}_1} }{ c_W^2 } \left( \frac{m^2_{\tilde{t}_1}}{m_Z^2} - 1 \right) 
	\left[ \frac{ \left( \cos^2 \theta_{\tilde{t}} -\frac{4}{3}s^2_W \right)^2 }{ 2m^2_{\tilde{t}_1} - m^2_Z }
	+ \frac{ \cos^2 \theta_{\tilde{t}} \sin^2 \theta_{\tilde{t}} }{m^2_{\tilde{t}_1} + m^2_{\tilde{t}_2} - m^2_Z } \right]& .
\end{align}

\subsubsection*{(6)  $\bm{ Z\gamma }$}

The squared matrix element is given by
\begin{equation}
\sum_{{\rm spin, \,helicity}} \left| \mathcal{M}
 \left( \tilde{t}_1 \tilde{t}^*_1 \rightarrow Z\gamma \right) \right|^2_{v=0} = 
	8 \left( \frac{2}{3} \right)^2 g^2 g'^2 \left( \frac{1}{2} \cos^2 \theta_{\tilde{t}} - \frac{2}{3} s_W^2 \right)^2,
\end{equation}
where $g^{'}$ is the hypercharge gauge coupling constant.

\subsubsection*{(7)  $\bm{ b \bar{b} }$}

The squared matrix element is given by
\begin{equation}
\sum_{\rm{color, \,spin}} \left| \mathcal{M} \left( \tilde{t}_1 \tilde{t}^*_1 \rightarrow b \bar{b} \right) \right|^2_{v=0} = 
	24 \left( m^2_{\tilde{t}_1} - m^2_b \right) \left( \frac{g \,m_b}{2m_W} \frac{\sin \alpha}{\cos \beta} \frac{c^{(2)}_{\tilde{t}_1}}{4m^2_{\tilde{t}_1} - m^2_h} \right)^2 .
\end{equation}

\subsubsection*{(8)  $\bm{ t \bar{t} }$}

The squared matrix element is given by
\begin{align}
\sum_{{\rm color, \,spin}}& \left| \mathcal{M}  
\left( \tilde{t}_1 \tilde{t}^*_1 \rightarrow t \bar{t} \right)
 \right|^2_{v=0} = 24 \left( m^2_{\tilde{t}_1} - m^2_t \right) \notag \\
	\times &\left\{ 
	\frac{1}{3} \frac{ m_{\tilde{\chi}^0_1} \left( a_1^2 - b^2_1 \right) + m_t  \left( a_1^2 + b^2_1 \right) }{
	m_t^2 - m_{\tilde{t}_1}^2 - m_{\tilde{\chi}^0_1}^2 }
	-  \frac{g \,m_t}{2 m_W} \frac{\cos \alpha}{\sin \beta} \frac{c^{(2)}_{\tilde{t}_1}}{4m^2_{\tilde{t}_1} - m^2_h} 
	\right\}^2 ,
\end{align}
where
\begin{eqnarray}
a_1 = -\frac{1}{\sqrt{2}} g' \left( \frac{1}{6} \cos \theta_{\tilde{t}} -\frac{2}{3} \sin \theta_{\tilde{t}} \right) , \\[+3pt]
b_1 = -\frac{1}{\sqrt{2}} g' \left( \frac{1}{6} \cos \theta_{\tilde{t}} +\frac{2}{3} \sin \theta_{\tilde{t}} \right) .
\end{eqnarray}

\subsubsection*{(9)  $\bm{ \tilde{\chi}^0_1 \tilde{\chi}^0_1 }$}

The squared matrix element is given by
\begin{equation}
\sum_{{\rm spin}} \left| \mathcal{M}
\left( \tilde{t}_1 \tilde{t}^*_1 \rightarrow  \tilde{\chi}^0_1 \tilde{\chi}^0_1 \right) \right|^2_{v=0} = 
	8 \left( m^2_{\tilde{t}_1} - m^2_{ \tilde{\chi}^0_1 } \right) 
	\left\{ \frac{ 2m_t \left( a_1^2-b_1^2 \right) + 2 m_{\tilde{\chi}^0_1} \left( a_1^2-b_1^2 \right) }
	{m_{\tilde{\chi}^0_1}^2 - m^2_{\tilde{t}_1} -m_t^2 }
	\right\}^2 .
\end{equation}

%%%%%%%%%%%%%%%%%%%%%%%%%%%%%%%%%%

%%%%%%%%%%%%%%%%%%%%%%%%%%%%%%%%%%%%%%%%%%%%%%%%%%%%%%%%

\end{document}